\documentclass[prd,preprint,superscriptaddress,showpacs,nofootinbib,%
tightenlines]{revtex4}
\usepackage{mathrsfs}
\usepackage{latexsym,bm}
\usepackage{CJK}
\usepackage{graphicx}
\usepackage{indentfirst}
\usepackage{slashed}
\usepackage{amsmath}
\usepackage{amssymb}
\usepackage{color}
\usepackage{hyperref}
\usepackage{epsfig}
\hypersetup{CJKbookmarks=true}
\usepackage[titletoc]{appendix}
\usepackage{multirow}%
\usepackage{rotating}
\usepackage{epstopdf}
\usepackage{extarrows}\usepackage{tabularx}
\usepackage{soul} \newcommand{\ha}{{h}}
\newcommand{\ga}{{g}}
\newcommand{\fp}{F_\pi}
\newcommand{\mpi}{M}
\newcommand{\mH}{\mathcal{H}}
\newcommand{\mHO}{\overset{\circ}{\mathcal{H}}}

\newcommand{\be}{\begin{equation}} \newcommand{\ee}{\end{equation}}
\newcommand{\ba}{\begin{array}{c}} \newcommand{\ea}{\end{array}}
\newcommand{\bea}{\begin{eqnarray}} \newcommand{\eea}{\end{eqnarray}}

\newcommand{\clm}{{m}}

\newcommand{\mn}{{m}}
\newcommand{\md}{{m}_\Delta}

\begin{document}
\title{%\vspace{-3cm}\hfill{\small\color{red} Version 9}\\[1em]\Large\bf %Manifestly Lorentz invariant
Pion-nucleon scattering %amplitude
in  covariant baryon chiral perturbation theory with explicit Delta resonances}
\author{De-Liang Yao}
\email{d.yao@fz-juelich.de}
\affiliation{Institute for Advanced Simulation, Institut f\"ur Kernphysik
   and J\"ulich Center for Hadron Physics, Forschungszentrum J\"ulich, D-52425 J\"ulich,
Germany}
\author{D.~Siemens}
\email{dmitrij.siemens@rub.de}
 \affiliation{Institut f\"ur Theoretische Physik II, Ruhr-Universit\"at Bochum,  D-44780 Bochum,
 Germany}
 \author{V.~Bernard}
 \email{bernard@ipno.in2p3.fr}
 \affiliation{Groupe de Physique Th\'eorique, Institut de Physique
Nucl\'eaire, UMR 8606,
CNRS,
Univ.~Paris-Sud, Universit\'e Paris-Saclay, 91405 Orsay Cedex, France}
 \author{E.~Epelbaum}
  \email{evgeny.epelbaum@ruhr-uni-bochum.de}
 \affiliation{Institut f\"ur Theoretische Physik II, Ruhr-Universit\"at Bochum,  D-44780 Bochum,
 Germany}
 \author{A.~M.~Gasparyan}
\email{ashot.gasparyan@ruhr-uni-bochum.de}
 \affiliation{Institut f\"ur Theoretische Physik II, Ruhr-Universit\"at Bochum,  D-44780 Bochum,
 Germany}
 \affiliation{SSC RF ITEP, Bolshaya Cheremushkinskaya 25, 117218 Moscow, Russia}
\author{J.~Gegelia}
\email{j.gegelia@fz-juelich.de}
\affiliation{Institute for Advanced Simulation, Institut f\"ur Kernphysik
   and J\"ulich Center for Hadron Physics, Forschungszentrum J\"ulich, D-52425 J\"ulich,
Germany}
\affiliation{Tbilisi State  University,  0186 Tbilisi,
 Georgia}
 \author{H.~Krebs}
 \email{hermann.krebs@rub.de}
 \affiliation{Institut f\"ur Theoretische Physik II, Ruhr-Universit\"at Bochum,  D-44780 Bochum,
 Germany}
\author{Ulf-G.~Mei\ss ner}
\email{meissner@hiskp.uni-bonn.de}
 \affiliation{Helmholtz Institut f\"ur Strahlen- und Kernphysik and Bethe
   Center for Theoretical Physics, Universit\"at Bonn, D-53115 Bonn, Germany}
 \affiliation{Institute for Advanced Simulation, Institut f\"ur Kernphysik
   and J\"ulich Center for Hadron Physics, Forschungszentrum J\"ulich, D-52425 J\"ulich,
Germany}

\begin{abstract}
We present the results of a third order calculation of the pion-nucleon scattering amplitude
in a chiral effective field theory with pions, nucleons and delta resonances as explicit
degrees of freedom. We work in a manifestly Lorentz invariant formulation of baryon chiral
perturbation theory  using dimensional regularization and the extended on-mass-shell
renormalization scheme. In the delta resonance sector, the on mass-shell renormalization
is realized as a complex-mass scheme. By fitting the low-energy constants  of the
effective Lagrangian to the $S$- and $P$-partial waves a satisfactory description of
the phase shifts  from the analysis of the Roy-Steiner equations is obtained. We predict
the phase shifts for the $D$ and $F$ waves and compare them with the results of the
analysis of the George Washington University group. The threshold parameters are calculated
both in the delta-less and delta-full cases. Based on the determined low-energy constants, 
we discuss the pion-nucleon sigma term.  Additionally, in order to determine the strangeness content
of the nucleon, we calculate the octet baryon masses in the presence of decuplet resonances
up to next-to-next-to-leading order in SU(3) baryon chiral perturbation theory. The octet 
baryon sigma terms are predicted as a byproduct of this calculation.

\end{abstract}
%\keywords{pion-nucleon scattering, baryon chiral perturbation theory, partial wave analysis}

\date{March 11, 2016}

\maketitle

\section{Introduction}
Elastic pion-nucleon ($\pi N$) scattering has been studied extensively since the middle of
the last century (see e.g. Refs.~\cite{Bransden:1973, Hohler:1983}), not only due to the wealth
of  experimental data, but also because of its importance for our understanding
of  chiral dynamics of quantum chromodynamics (QCD). From the theory side, in order to
describe such a fundamental process, dispersion relations for $\pi N$  scattering have been
investigated several decades ago~\cite{Chew:1957zz, Hamilton:1963zz,Steiner:1970mh}
and many phenomenological
models and different approaches have been proposed (see, e.g., Refs.~\cite{Olsson:1975th, Bofinger:1991ed, Gross:1992tj,Schutz:1998jx,Krehl:1999ak,Gasparyan:2003fp,Gasparyan:2010xz,Ronchen:2012eg}).
Roy-Steiner equations for the pion-nucleon scattering have been also analysed
recently~\cite{Ditsche:2012fv,Hoferichter:2015dsa,Hoferichter:2015tha,Hoferichter:2015hva}.
In the low-energy region a systematic and powerful tool to study  $\pi N$ scattering  is
provided by chiral perturbation theory (ChPT) ~\cite{Weinberg:1978kz, Gasser:1983yg, Gasser:1984gg,Scherer:2012xha}. An extension of the range of applicability of chiral effective field
theory (EFT) beyond the low-energy region has been also suggested in the recent work
of Ref.~\cite{Epelbaum:2015vea}.

	ChPT is the EFT of QCD in the low-energy region, which is widely used in modern
hadronic physics. It has the same symmetries as its underlying theory and is based on an
expansion in powers of small quark masses and external momenta, collectively denoted by $p$.
According to the power counting rules of ChPT, powers of $p$ are assigned to Feynman
diagrams and used to estimate the relative importance of their contributions in physical
quantities. Hence, low-energy physical quantities, which can not be obtained within 
perturbative QCD, are calculated in a perturbative expansion in powers of $p$.
Purely mesonic ChPT, the theory of Goldstone bosons only, has been very successful
\cite{Gasser:1983yg,Gasser:1984gg}. However, in baryon chiral perturbation theory (BChPT),
which additionally takes baryons into account,  the power counting becomes subtle due to
the non-zero baryon masses  in the chiral limit. A first attempt to study elastic $\pi N$
scattering using BChPT was made in Ref.~\cite{Gasser:1987rb}.  Therein, the power counting
rule was shown to be broken when the baryon propagators are involved in loop integrals,
namely loop diagrams give contributions of orders lower than assigned by the power counting.

To remedy this power counting breaking (PCB) issue, several approaches were proposed during
the last decades.  The most well-known approach is to calculate  physical quantities within
heavy baryon chiral perturbation theory (HBChPT)~\cite{Jenkins:1990jv, Bernard:1992qa}. In order
to restore the power counting,  a simultaneous expansion in $p$ and in inverse powers of the
baryon mass  is performed.  Within this framework  $\pi N$ scattering was analysed in detail
up to order ${\cal O}(p^3)$~\cite{Mojzis:1997tu, Fettes:1998ud} and  later up to order
${\cal O}(p^4)$~\cite{Fettes:2000xg}. A good description of partial wave phase shifts has
been achieved near  threshold.  However, the non-relativistic heavy baryon expansion 
distorts the analytic structure of the amplitudes, e.g. the location of the poles of baryon 
propagators are shifted, and also  convergence problems are encountered in
certain low-energy regions, e.g. for the scalar form factor of the nucleon at 
$t=4 M_\pi^2$~\cite{Becher:1999he}. It should be noted, however, that the proper analytic 
structure can be regained if one includes subleasing terms in the heavy baryon
propagator, see e.g.~\cite{Bernard:1993ry}. In any case, it is of  interest to treat the 
PCB problem within covariant BChPT. A pioneering work in Ref.~\cite{Ellis:1997kc} restored the 
power counting by keeping only the so-called soft parts of the Feynman diagrams. Successively, 
a much more elegant approach, known as infrared regularization (IR), was proposed in 
Ref.~\cite{Becher:1999he} and later extended/reformulated in Refs.~\cite{Goity:2001ny,Bernard:2003xf,Schindler:2003xv,Bruns:2008ub}. All the Feynman integrals in the IR regularization scheme are divided 
into infrared singular parts, respecting the power counting rules, and infrared regular parts, 
possibly violating them, and therefore the latter are discarded by means of absorbing them in 
(an infinite number of) the low-energy constants (LECs) of the effective Lagrangian. By making 
use of the IR scheme,  $\pi N$ scattering has been studied  up to ${\cal O}(p^4)$ order 
\cite{Becher:2001hv} (see also Ref.~\cite{Alarcon:2011kh} for ${\cal O}(p^3)$ order calculation). 
Besides, the analyses of the isospin violation and the SU(3) sector of BChPT have also been
considered in Refs.~\cite{Hoferichter:2009gn} and \cite{Mai:2009ce}, respectively. However, 
the IR regularization has its own drawbacks: the presence of an unphysical $u$-channel 
cut~\cite{ Becher:1999he, Ellis:1997kc} and the prediction of a large discrepancy of the 
Goldberger-Treiman (GT) relation~\cite{Alarcon:2011kh}. All these problems are due to dropping 
the entire infrared regular parts.

The extended-on-mass-shell (EOMS) scheme, developed in 
Refs.~\cite{Gegelia:1999gf,Gegelia:1999qt,Fuchs:2003qc}, is an alternative
approach to solve the PCB problem. It removes the power counting breaking terms (PCBT) at 
the level of amplitudes (or observables) by absorbing them into the renormalization of LECs 
of the effective Lagrangian. This is due to the fact that the PCBTs are polynomials of momenta
and/or quark masses.
  % It is the very essence of extended-on-mass-shell (EOMS) scheme developed in Refs.~\cite{Gegelia:1994zz, Fuchs:2003qc}.
 The EOMS scheme preserves the analytic structure of the physical quantities, e.g. scattering amplitudes.
%, by merely shifting the values of the LECs such that the power counting works again.
$\pi N$ scattering has been calculated using EOMS scheme in Ref.~\cite{Alarcon:2012kn} up 
to order ${\cal O}(p^3)$  and in
Ref.~\cite{Chen:2012nx}  up to order ${\cal O}(p^4)$. Contributions to the scattering amplitudes 
obtained  in those works possess  the correct power counting  and correct analytic properties.  
Moreover, the existing results of partial wave analysis are described well, and
remarkably, reasonable predictions for discrepancy of GT relation and the pion-nucleon sigma 
term $\sigma_{\pi N}$ are obtained. Nevertheless, as pointed out in Ref.~\cite{Alarcon:2012kn}, the 
convergence of the chiral expansion of the ${\cal O}(p^3)$ order $\pi N$
scattering amplitude within EOMS scheme is questionable when the $\Delta(1232)$ is not 
taken as an explicit degree of freedom.
This implies the necessity of including the $\Delta(1232)$ resonances as explicit degrees of 
freedom in the effective Lagrangian, together with nucleons and pions.

In this work, we present the full third order (leading one loop) calculation of the pion-nucleon 
scattering amplitude in a manifestly Lorentz invariant formulation of
BChPT with explicit deltas. We perform renormalization using the EOMS scheme such that the power 
counting violating terms are dealt with properly.
The $S$- and $P$-wave phase shifts are extracted from the manifestly Lorentz invariant amplitudes 
and then fitted to the phase shifts obtained from the recent Roy-Steiner (RS) equation analysis 
of $\pi N$ scattering~\cite{Hoferichter:2015hva} such that all involved LECs are determined. 
Based on the obtained LECs, we predict the $D$- and $F$-wave phase shifts and compare them with 
the results of the George Washington University (GWU) group analysis~\cite{SAID}.  The threshold 
parameters are determined for both the delta-less and delta-full cases. In addition, we discuss 
the pion-nucleon sigma term and the strangeness content of the nucleon in SU(3) BChPT.

This paper is organized as follows. In Section~\ref{one} we introduce the notations and the 
kinematics for the pion-nucleon scattering amplitudes.  Terms of the chiral effective Lagrangian 
that are needed for our third order calculation of the pion-nucleon scattering amplitude are 
specified in Section~\ref{effL}.  Contributions of tree and one loop diagrams in the scattering 
amplitude are discussed in Section~\ref{Acontr}. Renormalization of the one loop
diagrams and the definitions  of the pion-nucleon,  $g_{\pi N}$, and  pion-nucleon-delta, 
$g_{\pi N \Delta}$, couplings are given in Section~\ref{Renormalization}.  Section~\ref{PWPSs}  
contains the extraction of the phase shifts. The baryon sigma terms and the strangeness content 
of the nucleon are discussed in Section~\ref{sec:sigmaterms}.
We summarize our results in Section~ \ref{summary}  and the appendices contain explicit 
expressions of various quantities as well as other technicalities.

\section{Formal aspects of elastic pion-nuleon scattering}
\label{one}

\subsection{Kinematics and the structure of the amplitude}

The on-shell Lorentz- and time-reversal invariant  $T$-matrix for the elastic
scattering process $\pi^a(q)+N(p)\to\pi^{a^\prime}(q^\prime)+N(p^\prime)$, with Cartesian
isospin indices $a^\prime$ and $a$, depends on three Mandelstam variables
\bea
s=(p+q)^2\ ,\quad t=(p-p^\prime)^2\ ,\quad u=(p-q^\prime)\ ,
\eea
subject to the constraint
\bea
s+t+u=2 m_N^2+2M_\pi^2\, ,
\eea
with $m_N$ and $M_\pi$ the physical masses of the nucleon and the pion, respectively.
In the isospin limit, the amplitude $T^{a^\prime a}_{\pi N}(s,t,u)$ can be parameterized as
\bea
T_{\pi N}^{a^\prime a}(s,t,u)=\chi_{N^\prime}^\dagger\left\{\delta_{a^\prime a}T^+(s,t,u)
+\frac{1}{2}[\tau_{a^\prime},\tau_a]T^-(s,t,u)\right\}\chi_N\ ,
\eea
where $\tau_i$ are the Pauli matrices and $\chi_{N}$, $\chi_{N^\prime}$ denote nucleon iso-spinors.
Unless stated otherwise, the argument $u$ is always to be understood as a
function of $s$ and $t$, $u(s,t)=2 m_N^2+2M_\pi^2-s-t$.
The Lorentz decomposition of the invariant amplitudes $T^{\pm}$ reads
\bea\label{eq.DB}
T^{\pm}(s,t,u)=\bar{u}^{(s^\prime)}(p^\prime)\left\{D^\pm(s,t,u)
-\frac{1}{4m_N}[\slashed{q}^\prime,\slashed{q}]B^\pm(s,t,u)\right\}u^{(s)}(p)\ ,
\eea
with the superscript $(s^\prime)$, $(s)$ denoting the spins of the Dirac spinors
$\bar{u}$, $u$, respectively. The Lorentz decomposition is not unique, a popular alternative
form is
 \bea\label{eq.AB}
T^{\pm}(s,t,u)=\bar{u}^{(s^\prime)}(p^\prime)\left\{A^\pm(s,t,u)
+\frac{1}{2}\left(\slashed{q}^\prime+\slashed{q}\right)B^\pm(s,t,u)\right\}u^{(s)}(p)\ .
\eea
Furthermore, $A^\pm$ can be related to $B^\pm$ and $D^\pm$ via $A^\pm=D^\pm-\nu B^\pm$
with $\nu\equiv(s-u)/(4m_N)$.
Nevertheless, it is well known that the decomposition in
terms of $D$ and $B$, i.e. Eq.~(\ref{eq.DB}), is better
suited to perform the chiral expansion of the invariant amplitudes, while
there exists cancellations between power counting violating contributions from $A$ and $B$.

The whole $T$-matrix $T_{\pi N}^{a^\prime a}$ is symmetric under the so-called crossing
operation between the $s$- and $u$-channels, i.e. interchanging the incoming pion (nucleon)
and the outgoing pion (nucleon). As a result, due to the crossing symmetry the
invariant amplitudes $A$, $B$ and $D$ have the following properties:
\bea\label{eq:crossing}
D^\pm(s,t,u)&=&\pm D^\pm(u,t,s)\ ,\nonumber\\
B^\pm(s,t,u)&=&\mp B^\pm(u,t,s)\ ,\nonumber\\
A^\pm(s,t,u)&=&\pm A^\pm(u,t,s)\ .
\eea

\subsection{Partial wave projection and unitarity}
The amplitudes with definite isospin quantum number $I$ can be deduced via
\bea
\mathcal{A}^{I=\frac{1}{2}}=\mathcal{A}^{+}+2\mathcal{A}^-\ ,\quad
\mathcal{A}^{I=\frac{3}{2}}=\mathcal{A}^+-\mathcal{A}^-\ ,
\eea
where $\mathcal{A}\in\{A,B,D\}$. All possible elastic $\pi N$ scattering processes
are associated with the above specified two isospin amplitudes: $\mathcal{A}^{I=\frac{1}{2}}$
and $\mathcal{A}^{I=\frac{3}{2}}$. The partial wave projection of the isospin amplitudes
is given by
\bea
\mathcal{A}_\ell^I(s)=\int_{-1}^{+1}\mathcal{A}^I(s,t(s,z_s))\,P_\ell(z_s){\rm d}z_s\ ,
\qquad z_s\equiv\cos\theta\ ,
\eea
where $\theta$ is the scatting angle in the center-of-mass (CMS) frame and
$P_\ell(z_s)$ are  Legendre polynomials. Further, $t$ is regarded as a function
of $s$ and $z_s$, i. e.
\bea
t(s,z_s)&=&(z_s-1)\frac{\lambda(s,m_N^2,M_\pi^2)}{2s}\ ,
\eea
where $\lambda(a,b,c)\equiv a^2+b^2+c^2-2ab-2bc-2cd$ is the K\"all\'en function.
The physically relevant partial wave amplitudes $f_{\ell\pm}^I(s)$ can be constructed
from $\mathcal{A}_\ell^I(s)$ through
\bea
f_{\ell\pm}^I(s)=\frac{1}{16\pi\sqrt{s}}&\bigg\{&(E_p+m_N)\left[A_\ell^I(s)
+\left(\sqrt{s}-m_N\right)B_\ell^I(s)\right]\nonumber\\
&+&(E_p-m_N)\left[-A_{\ell\pm1}^I(s)+\left(\sqrt{s}+m_N\right)B_{\ell\pm1}^I(s)\right]\bigg\}\ ,
\eea
with $E_p=\frac{s+m_N^2-M_\pi^2}{2\sqrt{s}}$ and the subscript $\ell\pm$ is an
abbreviation for the total angular momentun $J=\ell\pm\frac{1}{2}$. One popular
notation for all the partial waves is the spectroscopic one, $L_{2I,2J}$, with
$L=S,P,D,F, \ldots$ (corresponding to $\ell=0,1,2,3,\ldots$). In general, below
the inelastic threshold, $f_{\ell\pm}^I(s)$ obeys the partial wave unitarity:
\bea
{\rm Im } f_{\ell\pm}^I(s)=q(s) \left|f_{\ell\pm}^I(s)\right|^2,\qquad 
\text{or }\qquad S_{\ell\pm}^I(s) \left(S_{\ell\pm}^I(s)\right)^\dagger=1\ ,
\eea
where
\bea
S_{\ell\pm}^I(s)\equiv 1+2iq(s)f_{\ell\pm}^I(s)
\eea
with $q(s)={\lambda(s,m_N^2,M_\pi^2)}/({2\sqrt{s}})$, the modulus of the three-momentum
in the CMS frame. A commonly used parametric form of the partial wave amplitudes is
 \bea\label{eq:Smatrix}
S_{\ell\pm}^I(s)&=&e^{2i\delta_{\ell\pm}^I(s)}\ ,\nonumber\\
f_{\ell\pm}^I(s)&=&q(s)^{-1}e^{i\delta_{\ell\pm}^I(s)}\sin\delta_{\ell\pm}^I(s)
= \frac{1}{2iq(s)}\left\{e^{2i\delta_{\ell\pm}^I(s)}-1\right\}.
 \eea
Here, the so-called partial wave phase shifts $\delta_{\ell\pm}^I(s)$ are real-valued functions
and they can be expressed as
\bea\label{eq:psunitary}
\delta_{\ell\pm}^I(s)={\rm Arg}\{f_{\ell\pm}^I(s)\}=\frac{1}{2} {\rm Arg} \{S_{\ell\pm}^I(s)\}\ .
\eea

\subsection{Extracting phase shifts from perturbative amplitudes\label{sec:pshift}}

\medskip

In chiral EFT,  the scattering amplitude $f(s)$ can be calculated perturbatively
up to certain order $\mathcal{O}(p^P)$
(for simplicity, all indices of the amplitudes are suppressed in this section),
\bea
f_{P}(s)=f^{(1)}(s)+f^{(2)}(s)+\ldots+f^{(P)}(s)~.
\eea
The full amplitude $f(s)=f_{P=\infty}(s)$ satisfies the partial wave unitarity
condition exactly, however,  $f_{P\neq\infty}(s)$ does not. 
The phase shifts can be calculated using
\bea\label{eq:psgeneralized}
\delta(s)={\rm Arctan} \left\{\frac{q(s)}{{\rm Re}\left[f_P(s)^{-1}\right]}\right\} \ .
\eea
This is equivalent to constructing a unitarized amplitude $f^U(s)$ corresponding to
$f_P(s)$ by setting
\bea\label{eq:psU}
{\rm Re}\left[f^U(s)\right]&=& \mathcal{N}\cdot{\rm Re}\left[f_P(s)\right],\nonumber\\
{\rm Im}\left[f^U(s)\right]&=& \mathcal{N}\cdot q(s)\left|f_P(s)\right|^2=\mathcal{N}
\cdot q(s)\left[\left({\rm Re}f_P(s)\right)^2+\left({\rm Im}f_P(s)\right)^2\right]\,
\eea
and then extracting the phase shifts by substituting the partial wave amplitudes 
corresponding to $ f^U(s)$ in Eq.~(\ref{eq:psunitary}). Here, $\mathcal{N}$ is given by 
the expression
\bea
\mathcal{N}=\left[ \frac{\left[{\rm Re}f_P(s)\right]^2}{|f_P(s)|^2}+|q(s)f_P(s)|^2\right]^{-1}\, .
\eea
For all partial waves except $P_{33}$ we use  the following expression
\bea\label{psold}
\delta(s)={\rm Arctan} \left\{{q(s)}{{\rm Re}\left[f_P(s)\right]}\right\}\,.
\eea
For the non-resonant partial waves the phase shifts given by  Eqs.~(\ref{psold}) and  
(\ref{eq:psgeneralized}) differ by higher order contributions only.

\section{Effective Lagrangian}
\label{effL}

The chiral effective Lagrangian relevant for our calculation of the pion-nucleon
scattering amplitude up to order ${\cal O}(p^3)$ can be written as
\bea
{\cal L}_{{\rm eff}}= \sum_{i=1}^{2}{\cal L}_{\pi\pi}^{(2i)}+\sum_{j=1}^3{\cal L}^{(j)}_{\pi N}
+\sum_{k=1}^2{\cal L}^{(k)}_{\pi\Delta}+\sum_{l=1}^3{\cal L}^{(l)}_{\pi N\Delta},\label{lageff}
\eea
where the superscripts in brackets correspond to the chiral orders. The first two
terms in Eq.~(\ref{lageff}) are sufficient to perform an analysis of the $\pi N$
scattering without explicit deltas.
For the case including deltas as explicit degrees of freedom, one also needs the
last two terms, which introduce interactions of deltas with pions and nucleons.

We start with the purely mesonic sector for which the required terms are given by
\cite{Gasser:1983yg,Bellucci:1994eb}
\bea {\cal
L}_{\pi\pi}^{(2)}&=&  \frac{F^2}{4}\mbox{Tr}(\partial_\mu U \partial^\mu U^\dagger)
+\frac{F^2 M^2}{4}\mbox{Tr}(U^\dagger+ U) ,\nonumber\\
{\cal L}_{\pi\pi}^{(4)}&=&\frac{1}{8}l_4\langle u^\mu
u_\mu\rangle\langle \chi_+\rangle+\frac{1}{16}(l_3+l_4)\langle
\chi_+\rangle^2, \eea
where $\langle ~~ \rangle$ denotes the trace in flavor space, $F$ is the pion
decay constant in the chiral limit, and $l_3,l_4$ are mesonic LECs.
The chiral operators, $u^\mu$ and $\chi_+$, are defined as
\bea\label{exp}
u_\mu&=&i\left[u^\dag(\partial_\mu-i r_\mu)u-u(\partial_\mu-i l_\mu)
u^\dag\right],\quad
U=u^ 2=\exp\left(\frac{i\,\tau^a \pi^a}{F}\right) ,
\nonumber\\
\chi^\pm&=&u^\dag\chi u^\dag\pm u\chi^\dag u~,\quad \chi= \left[ \begin{array}{c c}
      M^2 & 0 \\
      0 & M^2 \\
   \end{array}\right],
   \eea
with $M$ the leading order contribution to the charged pion mass.
Further,  the Goldstone bosons $\pi^ a$ are
incorporated in a $2\times2$ matrix-valued field $U$. In our present calculation
the external sources $l_\mu$ and $r_\mu$ can be set to zero, $l_\mu=r_\mu=0$.

Terms of the effective Lagrangian of the one-nucleon sector of
BChPT \cite{Fettes:1998ud} relevant for the $\pi N$ scattering are given as
\bea
{\cal L}_{\pi N}^{(1)}&=&\bar{\Psi}_N\left\{i\slashed{D}-m
+\frac{1}{2}g \,\slashed{u}\gamma^5\right\}\Psi_N\ ,\nonumber\\
{\cal L}_{\pi
N}^{(2)}&=&\bar{\Psi}_N\left\{c_1\langle\chi_+\rangle-\frac{c_2}{4m^2}\langle
u^\mu u^\nu\rangle(D_\mu D_\nu+h.c.)+\frac{c_3}{2}\langle
u^\mu
u_\mu\rangle-\frac{c_4}{4}\gamma^\mu\gamma^\nu[u_\mu,u_\nu]\right\}\Psi_N\ ,\nonumber\\
{\cal L}_{\pi N}^{(3)}&=&\bar{\Psi}_N\left\{-\frac{d_1+d_2}{4m}\big([u_\mu,[D_\nu,u^\mu]
+[D^\mu,u_\nu]]D^\nu+h.c.)\right.\nonumber\\
&&\qquad+\frac{d_3}{12m^3}([u_\mu,[D_\nu,u_\lambda]](D^\mu D^\nu
D^\lambda+sym.)+h.c.\big)+i\frac{d_5}{2m}([\chi_-,u_\mu]D^\mu+h.c.)\nonumber\\
&&\qquad+i\frac{d_{14}-d_{15}}{8m}(\sigma^{\mu\nu}\langle[D_\lambda,u_\mu]u_\nu
-u_\mu[D_\nu,u_\lambda]\rangle
D^\lambda+h.c.)+\frac{d_{16}}{2}\gamma^\mu\gamma^5\langle\chi_+\rangle
u_\mu\nonumber\\
&&\qquad\left.+\frac{id_{18}}{2}\gamma^\mu\gamma^5[D_\mu,\chi_-]\right \}\Psi_N\ .
\eea
Here, $m$ and  $g$ denote the nucleon bare mass and the bare axial coupling constant,
respectively. The notion 'bare' will be explained below.
The LECs $c_i$ and $d_j$ have dimension GeV$^{-1}$ and GeV$^{-2}$, respectively.
The covariant derivative acting on the nucleon field is defined as
$D_\mu\Psi_N=(\partial_\mu+\Gamma_\mu)\Psi_N$ with
\bea\label{eq.gamma}
\Gamma_\mu=\frac{1}{2}\left\{u^\dagger(\partial_\mu-i\,r_\mu)u
+u(\partial_\mu-i\,l_\mu)u^\dagger\right\}\,.
\eea

Fields with spin-3/2 corresponding to the delta resonances can be described via the
Rarita-Schwinger formalism, where the field is represented by a
vector spinor $\Psi^{\mu}$ \cite{Rarita:1941mf}.
For the purposes of our calculation the following Lagrangians are needed \cite{Hemmert:1997ye},
\bea
{\cal L}^{(1)}_{\pi\Delta}&=&-\bar{\Psi}_{\mu}^i\xi^{\frac{3}{2}}_{ij}
\left\{\left(i\slashed{D}^{jk}-m_\Delta\delta^{jk}\right)g^{\mu\nu}
+iA\left(\gamma^\mu D^{\nu,jk}+\gamma^\nu D^{\mu,jk}\right)\right. \nonumber\\
&&+\frac{i}{2}(3A^2+2A+1)\gamma^\mu\slashed{D}^{jk}\gamma^\nu+m_\Delta
\delta^{jk}(3A^2+3A+1)\gamma^{\mu}\gamma^\nu
\nonumber\\
 &&\left.+\frac{g_1}{2}\slashed{u}^{jk}\gamma_5g^{\mu\nu}+\frac{g_2}{2}
(\gamma^\mu u^{\nu,jk}+u^{\nu,jk}\gamma^\mu)\gamma_5
+\frac{g_3}{2}\gamma^\mu\slashed{u}^{jk}\gamma_5\gamma^\nu \right\}
\xi^{\frac{3}{2}}_{kl}{\Psi}_\nu^l\  ,
 \label{Dlagr0}
 \\
{\cal L}^{(2)}_{\pi\Delta}&=&a_1 \bar{\Psi}_{\mu}^i\xi^{\frac{3}{2}}_{ij}
\Theta^{\mu\alpha}(z)\langle\chi_+\rangle\delta^{jk}
g_{\alpha\beta}\Theta^{\beta\nu}(z^\prime)\xi_{kl}^{\frac{3}{2}}\Psi_\nu^l\ ,
\label{DLagr}
\eea
where $m_\Delta$ and $g_1$, $g_2$, $g_3$, $a_1$ are the bare mass of the delta and
bare coupling constants, respectively. Further, $\Theta^{\mu\alpha}=g^{\mu\alpha}
+z\gamma^\mu\gamma^\nu$, where $z$ is a so-called off-shell parameter.

%It is redundant in the sense that the physical observables do not depend on it.
The isospin-$\frac{3}{2}$ projection operator is defined as
$\xi^{\frac{3}{2}}_{ij}=\delta_{ij}-\tau_i\tau_j/3$.
${\Psi}_\mu^i$ is a short-hand notation for ${\Psi}_{\mu,\alpha,i,r}$, which is a
vector-spinor isovector-isospinor field, with $\mu$ being a Lorentz vector index,
$\alpha$ Dirac spinor index, $i$ an isovector index, and $r$ an isospinor index.
From now on, the Dirac spinor and the isospinor indices will be suppressed for
simplicity. The fields ${\Psi}_{\mu}^i$  are related to the physical $\Delta(1232)$
states $\Delta^{++}$, $\Delta^{+}$, $\Delta^{0}$ and $\Delta^{-}$ by
\bea
\xi_{1j}^{\frac{3}{2}}\Psi_{\mu}^j&=&\frac{1}{\sqrt{2}}\left(
                                     \begin{array}{c}
                                       \frac{1}{\sqrt{3}}\Delta_\mu^0-\Delta_\mu^{++} \\
                                       \Delta_\mu^--\frac{1}{\sqrt{3}}\Delta_\mu^+ \\
                                     \end{array}
                                   \right)\ ,\nonumber\\
\xi_{2j}^{\frac{3}{2}}\Psi_{\mu}^j&=&-\frac{i}{\sqrt{2}}\left(
                                     \begin{array}{c}
                                       \frac{1}{\sqrt{3}}\Delta_\mu^0+\Delta_\mu^{++} \\
                                       \Delta_\mu^-+\frac{1}{\sqrt{3}}\Delta_\mu^+ \\
                                     \end{array}
                                   \right)\ ,\nonumber\\
\xi_{3j}^{\frac{3}{2}}\Psi_{\mu}^j&=&\sqrt{\frac{2}{3}}\left(
                                     \begin{array}{c}
                                       \Delta_\mu^{+} \\
                                       \Delta_\mu^0 \\
                                     \end{array}
                                   \right)\ .
\eea
The covariant derivative acting on the Rarita-Schiwinger fields is defined as
\bea
{\cal D}_{\mu,ij}\Psi_{\nu}^j&=&\left(\partial_\mu\delta_{ij}-2i\epsilon_{ijk}\Gamma_{\mu,k}
+\delta_{ij}\Gamma_{\mu}\right)\Psi_{\nu}^j\ ,\nonumber\\
\Gamma_{\mu,k}&=&\frac{1}{2}\langle\tau_k\Gamma_\mu\rangle
=-\frac{i}{4F^2}\epsilon_{ijk}(\partial_\mu\pi^i)\pi^j
+\frac{i}{48F^4} \pi ^ a \pi^ a\epsilon_{ijk}(\partial_\mu\pi^i)\pi^j+{\cal O}(\pi^4),
\eea
with $\Gamma_\mu$ given by Eq.~(\ref{eq.gamma}).

Finally, the pion-nucleon-delta interaction part has the form
\bea\label{lag:piND}
{\cal L}^{(1)}_{\pi N\Delta}&=&h\,\bar{\Psi}_{\mu}^i\xi_{ij}^{\frac{3}{2}}\Theta^{\mu\alpha}(z_1)\ 
\omega_{\alpha}^j\Psi_N+h.c.\ ,\nonumber\\
{\cal L}^{(2)}_{\pi N\Delta}&=&\bar{\Psi}_{\mu}^i\xi_{ij}^{\frac{3}{2}}\Theta^{\mu\alpha}(z_2)
\left[i\,b_3\omega_{\alpha\beta}^j\gamma^\beta+i\,\frac{b_8}{m}\omega_{\alpha\beta}^ji\,D^\beta\right]
\Psi_N+h.c.\ ,\nonumber\\
{\cal L}^{(3)}_{\pi N\Delta}&=&\bar{\Psi}_{\mu}^i\xi_{ij}^{\frac{3}{2}}\Theta^{\mu\nu}(z_3)\left[
\frac{f_1}{m}[D_\nu,\omega_{\alpha\beta}^j]\gamma^\alpha i\,D^\beta-\frac{f_2}{2m^2}[D_\nu,\omega_{\alpha\beta}^j]
\{D^\alpha,D^\beta\}\right.\nonumber\\
&&\hspace{2.5cm}\left.+f_4\omega_\nu^j\langle\chi_+\rangle+f_5[D_\nu,i\chi_-^j]\right]\Psi_N+h.c.,
\eea
where the bare pion-nucleon-delta coupling constant at lowest order is denoted by
$h$ and $b_3$, $b_8$, $f_1$, $f_2$, $f_4$ and $f_5$ are  bare LECs of higher orders.
New off-shell parameters $z_1$, $z_2$ and $z_3$ appear in the interaction terms.
As discussed later,  they do not contribute in physical quantities.
For convenience, the three chiral structures, $\omega_\alpha^i$, $\omega_{\alpha\beta}^j$
and $\chi_-^k$, are introduced as building blocks of the Lagrangian.
Their explicit expressions are given by
\bea
\omega_\alpha^i&=&\frac{1}{2}\langle\tau^i u_\alpha\rangle=-\frac{1}{F}\partial_\alpha\pi^i
+\frac{1}{6F^3}(\partial_\alpha\pi^i \pi^ a \pi^ a-\pi^i\partial_\alpha \pi^ a\pi^a)
+{\cal O}(\pi^5)\ ,\nonumber\\
\omega_{\alpha\beta}^j&=&\frac{1}{2}\langle\tau^j[\partial_\alpha,u_\beta]\rangle
= -\frac{1}{F}\partial_\alpha\partial_\beta\pi^j+{\cal O}(\pi^3)\ ,\nonumber\\
\chi_-^k&=&\frac{1}{2}\langle\tau^k \chi_-\rangle=-\frac{2i}{F}M^2\pi^k+{\cal O}(\pi^3)\ ,
\eea
where we expanded  them in powers of pion fields to the order needed for our calculations.

%\section{Contributions to the amplitude}
\section{Calculation of the pion-nucleon amplitude up to NNLO}
\label{Acontr}

\subsection{Power counting}

For diagrams involving only pion and nucleon lines, we use the standard power counting
of Refs.~\cite{Weinberg:1991um,Ecker:1994gg}.
For diagrams with delta lines we apply the power counting of
Refs.~\cite{Hemmert:1996xg,Hemmert:1997ye}, that is we count the mass
difference  $\Delta = m_\Delta-m_N$ as of order ${\cal O}(p)$, although we do not
expand the interaction terms of the effective Lagrangian and the physical
quantities in $\Delta$. The above power counting leads to the dressing of the
delta propagator in the resonant region (for a different point of view,
see Refs.~\cite{Pascalutsa:2002pi,Hagelstein:2015egb}).

In particular, it is self-consistent to count $A-B\sim p^n$  if $A\sim p^n$ and
$B\sim p^n$, however, more care has to be taken when dealing with inverse powers of
similar differences. From $A\sim p^n$, $B\sim p^n$ it does {\it not} necessarily
follow that $1/(A-B)\sim p^{-n}$. For example, if we have $ A=M_\pi+ a M_\pi^3$ ($a\neq 0$)
and $B=M_\pi + b M_\pi^4 $ ($b\neq a$), by counting $A-B$ as of order $p$ we overestimate
this difference (which causes no problems), however, if we count $1/(A-B)$ as of
order $1/p$, we underestimate this quantity, which is apparently of order $1/p^3$
and that leads to inconsistency. Considering the delta propagator appearing in the
intermediate states in the $s$-channel diagrams
$D_0^{\mu\nu}\sim 1/(s-m_\Delta^2) =1/(s-m_N^2-2 m_N\Delta+{\cal O}(\Delta^2))$,
we count $s-m_N^2\sim p$, $\Delta\sim p$, however, it would be wrong to conclude
that $\sim 1/(s-m_\Delta^2) =1/(s-m_N^2-\Delta (m_N+m_\Delta))\sim 1/p$.
For $s\to m_\Delta^2$ this propagator diverges, so do all diagrams with multiple
self-energy insertions, therefore we need to sum up these diagrams, i.e. consider the
dressed propagator $D^{\mu\nu}(k) \sim 1/(\slashed k -m_\Delta-\Sigma(k) )$.
For $\slashed k\to m_\Delta$ we obtain $D^{\mu\nu}(k) \sim 1/(-\Sigma(k) )\sim 1/p^3$
as the leading contribution in the self-energy is of order ${\cal O}(p^3)$.
We follow an alternative way of dealing with the problem by using the complex-mass
scheme, specified later, where the undressed propagator contains the width of
the unstable particle and therefore the re-summation is not needed.

\subsection{Tree level contributions}

\begin{figure}[t]
\begin{center}
\epsfig{file=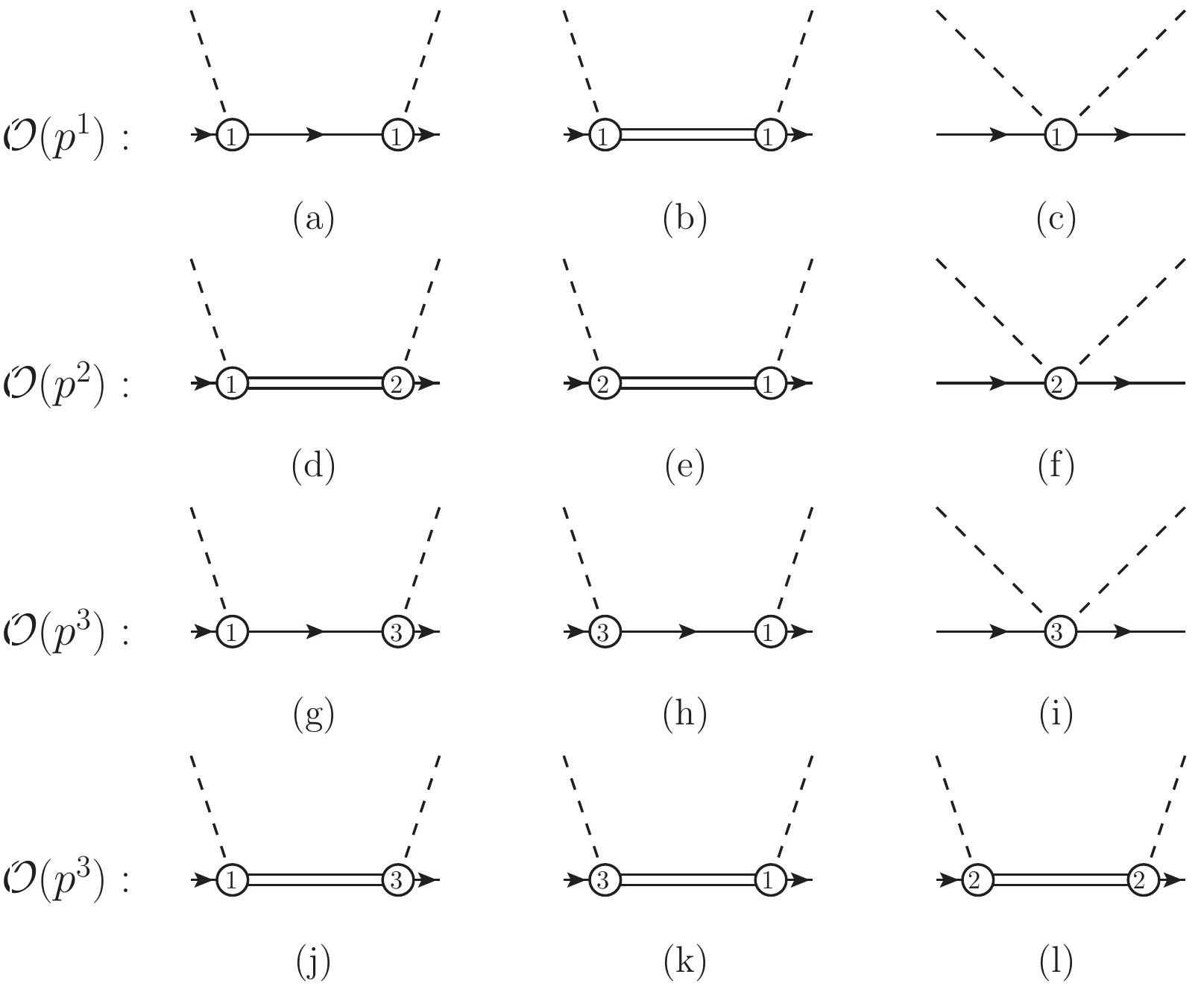,scale=0.6}
\caption{Tree level diagrams contributing to $\pi N$ scattering up to order
${\cal O}(p^3)$. Dashed, solid and double lines represent pions, nucleons and deltas,
respectively. Circled numbers mark the chiral orders of the vertices. Crossed
diagrams are not shown.  The diagrams in the first, second, third and fourth rows
are of $O(p^1)$, $O(p^2)$, $O(p^3)$ and $O(p^3)$, respectively.  }
\label{fig:tree}
\end{center}
\end{figure}

The  Feynman tree diagrams contributing to the pion-nucleon scattering amplitude up
to order ${\cal O}(p^3)$ are displayed in Fig.~\ref{fig:tree} with chiral orders
specified in front of them. The crossed diagrams are not shown since their
contributions can be obtained by using the crossing relations given in
Eq.~(\ref{eq:crossing}). The diagrams with mass insertions in propagators, which
are generated by the $c_1$ term in $\mathcal{L}_{\pi N}^{(2)}$ for the nucleon and
$a_1$ term in $\mathcal{L}_{\pi \Delta}^{(2)}$ for the delta, are not shown either.
Their contributions are automatically taken into account if one replaces the masses
in the nucleon and delta propagators by
\bea
m &\to&  m_2= m-4c_1 M^2\ ,\nonumber\\
m_\Delta &\to& m_{\Delta,2}= m_\Delta -4 a_1 M^2\ .
\eea
As can be seen from Figure~\ref{fig:tree}, there are three different types of
contributions: nucleon-exchange, contact-interaction and delta-exchange diagrams.
The $s$-channel Born-term contributions of the nucleon-exchange diagrams, namely the
sum of contributions of diagrams (a), (g) and (h), can be written as
\bea
D^\pm_N(s,t)&=&-\frac{g_2^2}{4F^2}\frac{2m_N}{s-m_2^2}
\left\{(s-m_N^2)(m_2+m_N)-\frac{s-u}{4m_N}\left(s+2m_Nm_2+m_N^2\right)\right\}\ ,\nonumber\\
B^\pm_N(s,t)&=&-\frac{g_2^2}{4F^2}\frac{s+2m_Nm_2+m_N^2}{s-m_2^2}\ ,
\eea
where $g_2\equiv g+2(2d_{16}-d_{18})M^2$. Here the appearance of $m_2$ is due to
the inclusion of the mass-insertion diagrams to the order  we are working.
The contact-term contributions, which are represented by Diagrams (c), (f) and (i), read
\bea\label{eq:contact}
D^+_C(s,t)&=&-\frac{4c_1 M^2}{F^2}+\frac{c_2(16m_N^2\nu^2-t^2)}{8F^2m^2}
+\frac{c_3(2M_\pi^2-t)}{F^2}\, ,\nonumber\\
D^-_C(s,t)&=&\frac{\nu}{2F^2}+\frac{2\nu}{F^2}\left\{2(d_1+d_2+2d_5)M_\pi^2
-(d_1+d_2)t+2d_3\nu^2\right\} ,\nonumber\\
B^+_C(s,t)&=&\frac{4(d_{14}-d_{15})\nu\, m_N}{F^2}\, ,\quad
B^-_C(s,t)=\frac{1}{2F^2}+\frac{2c_4m_N}{F^2}\, .
\eea

The calculation of the delta-exchange diagrams is performed using the Lagrangian
of Eq.~(\ref{lag:piND}), where the off-shell parameters $z_1$, $z_2$ and $z_3$ are
involved. As argued in Refs.~\cite{Tang:1996sq,Pascalutsa:2000kd,Krebs:2009bf},
those parameters are redundant in the sense that their contributions in physical
quantities can be absorbed into LECs of other interaction terms. The same applies
also to the $g_2$ and $g_3$ terms in the Lagrangian of Eq.~(\ref{Dlagr0}), therefore
for the convenience we take
\bea
g_2=g_3=z_1=z_2=z_3=0\ .
\eea
With the above specifications, the LO Born-term contribution of the
$\Delta$-exchange is
\bea\label{eq:delta}
D^+_\Delta(s,t)&=&\frac{h^2}{9F^2m_\Delta^3(m_\Delta^2-s)}\bigg\{\mathcal{F}_A(s,t)
-\frac{s-u}{4m_N}\mathcal{F}_B(s,t)\bigg\}\, ,\nonumber\\
B^+_\Delta(s,t)&=&-\frac{h^2}{9F^2m_\Delta^3(m_\Delta^2-s)}\mathcal{F}_B(s,t)\, ,\nonumber\\
D^-_\Delta(s,t)&=&-\frac{1}{2}D^+_\Delta(s,t)\ ,\qquad B^-_\Delta(s,t)=-\frac{1}{2}B^+_\Delta(s,t).
\eea
The definition of the functions $\mathcal{F}_{A,B}(s,t)$ is given in
Appendix~\ref{app:treedelta}. Similarly to the nucleon case, $m_\Delta$ should
be understood as $m_{\Delta,2}$ but we keep using $m_\Delta$ for short.
Tree order amplitudes corresponding to diagrams (d), (e), (j), (k) and (l)
are given in Appendix~\ref{app:treedelta}. However, they are redundant in the
sense that their contributions can be taken into account by the redefinition of
$h$ in Eq.~(\ref{eq:delta}) and the LECs in the contact terms, Eq.~(\ref{eq:contact}).
By redefining the $\pi N\Delta$ coupling $h$ as
\bea\label{eq:redefinition.h}
h&\to& h+\left(b_3\Delta_{23}+b_8\,\Delta_{123}\right)
+\left(f_1\Delta_{23}+f_2\,\Delta_{123}\right)\Delta_{123}-2(2f_4-f_5)M^2\ ,
\eea
with
\bea
\Delta_{123}\equiv \frac{M^2+m^2-m_\Delta^2}{2 m}\ ,\qquad\Delta_{23}\equiv m-m_\Delta\ ,
\eea
the pole structures in the $\mathcal{O}(p^2)$ and $\mathcal{O}(p^3)$ order
delta-exchange diagrams are absorbed. Further, the remaining none-pole parts
can be absorbed by making use of
\bea\label{eq:redefinition.cd}
c_i&\to& c_i+\delta c_i\ , \qquad (i=1,2,3,4)\ ,\nonumber\\
d_j&\to& d_j+ \delta d_j\ ,\qquad(j=1,2,3,5,14,15).
\eea
The explicit expressions for $\delta c_i$ and $\delta d_j$ are given in
Appendix~\ref{app:treedelta}.

Finally,  if the redefinitions of Eqs.~(\ref{eq:redefinition.h})
and (\ref{eq:redefinition.cd}) are imposed, the tree contribution can
be summarized as
\bea
D_{\rm tree}^\pm(s,t)&=&D^\pm_N(s,t)\pm D^\pm_N(u,t)
+D^\pm_\Delta(s,t)\pm D^\pm_\Delta(u,t)+D^\pm_C(s,t)\ ,\nonumber\\
B_{\rm tree}^\pm(s,t)&=&B^\pm_N(s,t)\mp B^\pm_N(u,t)
+B^\pm_\Delta(s,t)\mp B^\pm_\Delta(u,t)+B^\pm_C(s,t)\ ,
\eea
where Eq.~(\ref{eq:crossing}) has been used.

\subsection{Leading one-loop contributions}

\begin{figure}[t]
\begin{center}
\epsfig{file=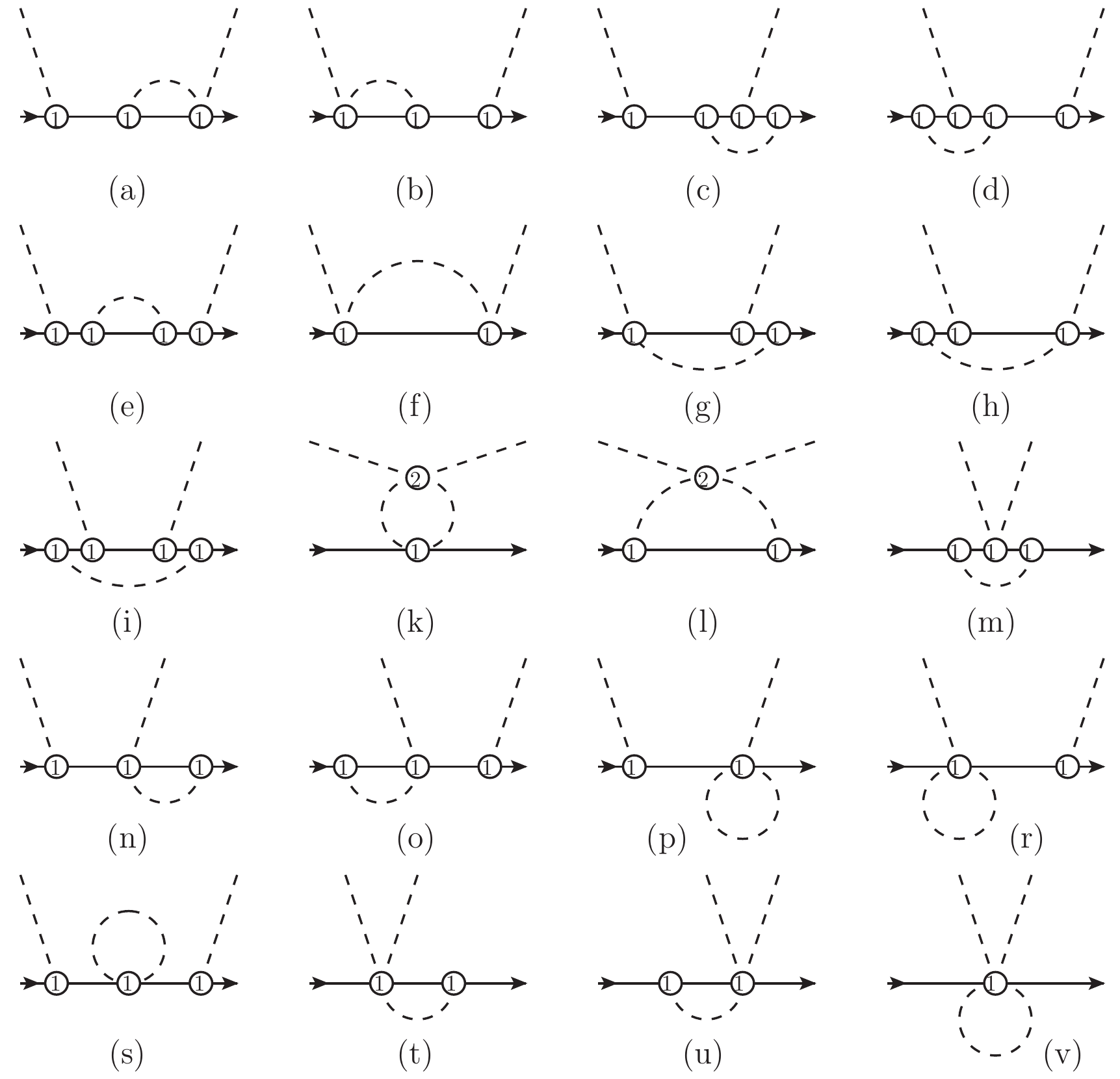,scale=0.6}
\caption{One-loop Feynman diagrams without explicit deltas to order $O(p^3)$.
Dashed and solid  lines represent pions and nucleons, respectively.
Circled numbers mark the chiral orders of the vertices. Crossed diagrams  are not shown.}
\label{fig:loop}
\end{center}
\end{figure}

\begin{figure}[t]
\begin{center}
\epsfig{file=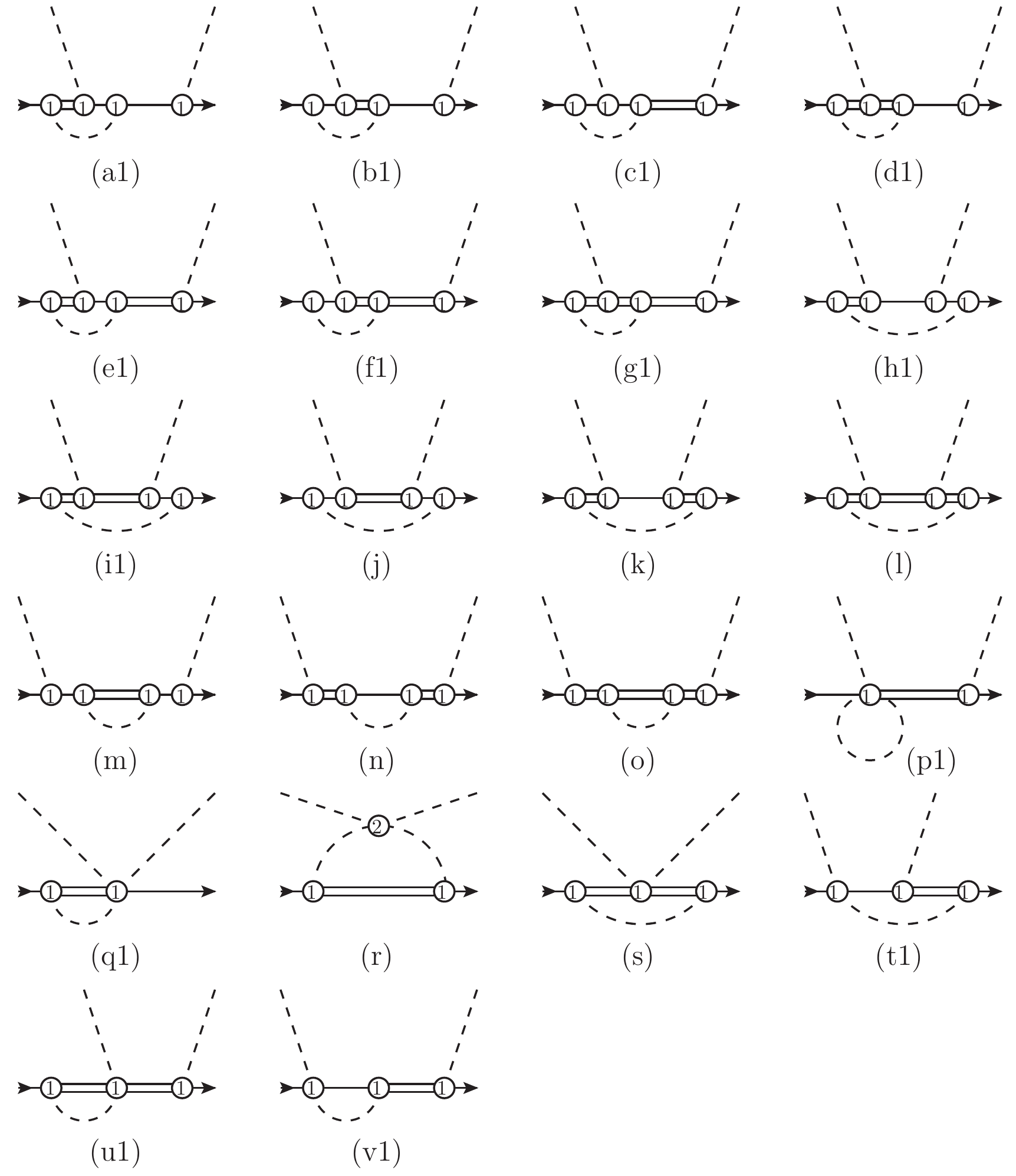,scale=0.6}
\caption{One-loop Feynman diagrams with explicit deltas  to  order $O(p^3)$.
Dashed, solid and double lines represent pions, nucleons and deltas, respectively.
Circled numbers mark the chiral orders of the vertices. Crossed diagrams, diagrams
with the reversed time ordering and  diagrams giving vanishing contributions
are not shown.}
\label{fig:loopD}
\end{center}
\end{figure}

The one-loop Feynman diagrams up to order  $\mathcal{O}(p^3)$, without and
with explicit deltas, are shown in Fig.~\ref{fig:loop} and Fig.~\ref{fig:loopD},
respectively. For easier comparison, the labeling scheme for the delta-less
loop diagrams of Refs.~\cite{Alarcon:2012kn,Chen:2012nx} is followed.
For the explicit expressions of the contributions
of the delta-less one-loop diagrams in the amplitudes we refer the reader to
Refs.~\cite{Alarcon:2012kn,Chen:2012nx}. We have reproduced their results.

Due to the complexity of the spin-$3/2$ delta propagator, the calculation of
the delta-full loop diagrams in Fig.~\ref{fig:loopD} is much more complicated.
All diagrams have been calculated using two independent computer codes giving
identical results. The final expressions are much too huge to be displayed in the
paper, however, they can be obtained from the authors.

\section{\label{Renormalization}Renormalization}
%and definition of coupling constants}

To calculate the loop diagrams we apply  dimensional regularization with $d$
the number of space-time dimensions. All UV divergences of physical quantities are
removed by counter terms generated by the effective Lagrangian and absorbed in the
corresponding LECs. The finite pieces of the subtraction terms are fixed such
that the subtracted contributions of the loop diagrams in physical quantities
satisfy the  power counting.
%{\color{red}\st{While we do not show the counter terms explicitly in the effective Lagrangian the fields and parameters of the effective Lagrangian are to be interpreted as renormalized %quantities. }}
 The required counter terms are generated by splitting the bare
parameters as follows:
\bea
X&\equiv&X_R+\frac{\bar{\delta}X}{16\pi^2F^2}R+\frac{\bar{\bar\delta}X}{16\pi^2F^2}\ ,
\qquad X\in\{g,h,m,m_\Delta,a_1,c_{i=1,\cdot 4}\} \ ,\label{eq.EOMSshifts}\\
Y&\equiv&Y_R+\frac{\bar{\delta}Y}{16\pi^2F^2}R\ ,
\quad Y\in\{\ell_3,\ell_4,d_{1}+d_2,d_3,d_5,d_{14}-d_{15},d_{18}-2d_{16}\} \ ,\label{eq.MSbarshifts}
\eea
where $X_R$ and $Y_R$ are renormalized parameters and
$R\equiv {2}/({d-4})+\gamma_E-1-\ln(4\pi)$, and $\gamma_E$ is the Euler number. \footnote{Notice that to simplify notations below in tables with numerical values of renormalized LECs we suppress "R" subscripts.}

 Below we first introduce the renormalized and physical masses and wave function
renormalization constants followed by the definitions of the pion decay constant
$F_\pi$, the LO $\pi NN$ coupling $g_{\pi N}$ and LO $\pi N\Delta$ coupling $g_{\pi N\Delta}$.
In the calculations of $g_{\pi N}$ and $g_{\pi N\Delta}$ we follow a procedure analogous
that of Ref.~\cite{Schindler:2006it}. The $\pi N$ scattering amplitudes obtained
by using the EOMS scheme are discussed in the end.

\subsection{Masses and wave function renormalization constants}
\subsubsection{Pion}
The pions, nucleons and deltas are explicit degrees of freedom in our calculation.
Expressions for the pion wave function renormalization constant $\mathcal{Z}_\pi$ and
the pion pole mass $M_\pi^2$ at one-loop order have the form
(see e.g. Ref.~\cite{Scherer:2002tk})
\bea
\label{PWFRC}
\mathcal{Z}_\pi&=&1-\frac{1}{F^2}\left[2\ell_4 M^2+\frac{2}{3}\mH_\pi\right],\\
\label{eq.mpi}
M_\pi^2&=&M^2\left[1+\ell_3\frac{M^2}{F^2}-\frac{1}{2F^2}\mH_\pi\right],
\eea
where $\mH_\pi$ is a one-loop integral defined in the appendix together with
all loop integrals which contribute in our calculations. Since in dimensional
regularization there are no power counting violating terms from the loops,
the renormalization of the pion mass can be treated in the standard way of
mesonic ChPT by taking $\ell_3\equiv\ell_{3R}-\frac{3}{4}\frac{R}{16\pi^2}$.

The case of the nucleon is more complicated and can  be done in the EOMS scheme
so that the PCBTs from loops are dealt properly. We also give the renormalization
of the delta mass and the corresponding wave function renormalization constant.

\subsubsection{Nucleon\label{sec:nucleon}}

Defining $ -i \Sigma_N $ as the sum of one-particle irreducible diagrams contributing
to the nucleon two-point function, the dressed propagator of the nucleon is given as
\begin{equation}
i \,S_N(p) = \frac{i}{\slashed p-m-\Sigma_N(\slashed p)}
= \frac{i \mathcal{Z}_N}{\slashed p-m_N}+{\rm NP},
\label{NDProp}
\end{equation}
where $m_N$ is the nucleon pole mass  and $\mathcal{Z}_N$ is the wave function
renormalization constant. NP stands for the non-pole part (also for the delta
propagator below).

\begin{figure}[t]
\begin{center}
\epsfig{file=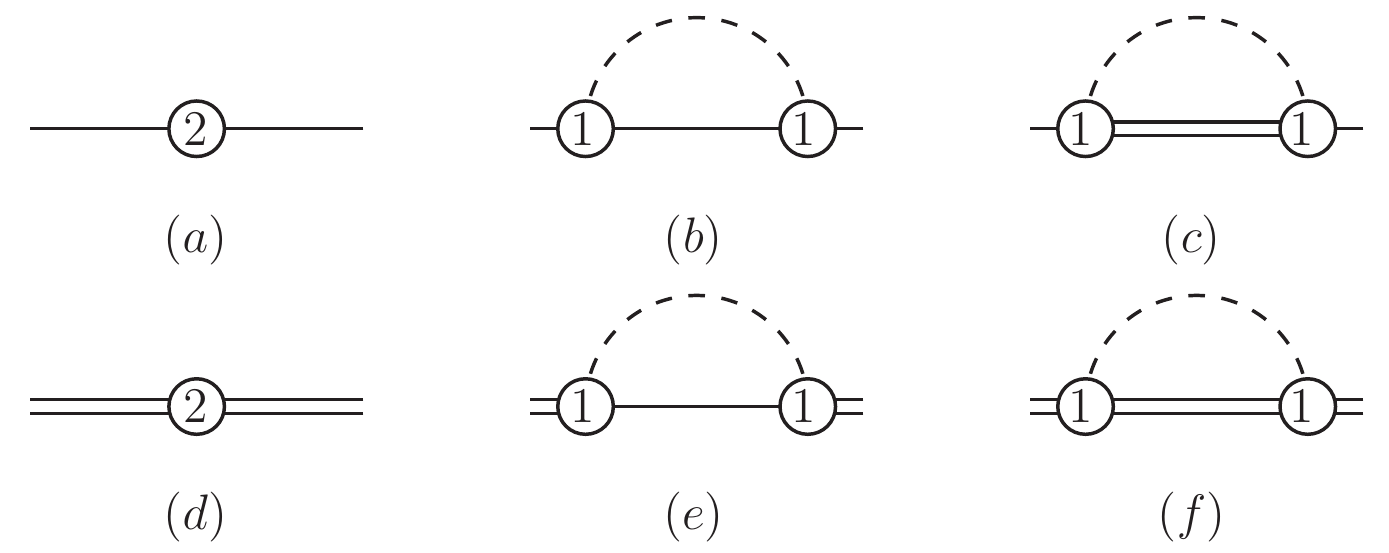,scale=0.6}
\caption{Tree and one-loop diagrams contributing to the self-energies of the nucleon
and the delta resonance up to the order $O(p^3)$.}
\label{fig:m}
\end{center}
\end{figure}

The nucleon self-energy up to order ${\cal O}(p^3)$ consists of tree and loop diagrams
shown in Fig.~\ref{fig:m} (a), (b) and (c) and the corresponding expressions are
given by 
\begin{equation}
 \Sigma_N(\slashed{p})
= -4c_1M^2+\Sigma_{A}^{N,{\rm loop}}(s)+\slashed{p}\,\Sigma_{B}^{N,{\rm loop}}(s)\ ,
\label{NSE}
\end{equation}
where $s\equiv p^2$ and the first term is the tree-order contribution.
The explicit expressions for $\Sigma_A^{N,{\rm loop}}(s)$ and $\Sigma_B^{N,{\rm loop}}(s)$ 
are shown in Appendix~\ref{sec.SE}.  The pole mass of the nucleon, $m_N$,
corresponding to Eq.~(\ref{NSE}) is given as the solution to the equation
\bea\label{eq.mN}
m_N=m-4c_1M^2+\Sigma_A^{N,{\rm loop}}(m_N^2)+m_N\,\Sigma_B^{N,{\rm loop}}(m_N^2)\,.
\eea
The nucleon wave function renormalization constant $\mathcal{Z}_N$ is given by
\bea
\mathcal{Z}_N&=&\left[\Sigma_B^{N,{\rm loop}}(s)+2s\frac{\partial}{\partial s}
\Sigma_B^{N,{\rm loop}}(s) +2\slashed{p}\frac{\partial}{\partial s}
\Sigma_A^{N,{\rm loop}}(s)\right]_{\slashed{p}=m_N} \, .
\eea
Within the on mass-shell renormalization  the renormalized mass of a particle
is chosen equal to the pole position of the corresponding dressed propagator. In case
of BChPT, where we  want to keep track of the quark mass dependence of physical
quantities explicitly, we use the EOMS scheme \cite{Fuchs:2003qc}, that is we
choose the renormalized mass of the nucleon as the pole mass in the chiral limit.
In order to cancel the UV divergence and PCBTs from the loop diagrams, one needs
to split the bare parameters $m$ and $c_1$ in Eq.~(\ref{eq.mN}) as specified by
Eq.~(\ref{eq.EOMSshifts}). The explicit expressions of the counter terms
${\bar\delta} m$, $\bar{\bar \delta}{m}$, ${\bar\delta}{c_1}$ and
$\bar{\bar\delta}{ c_1}$ are given in the Appendix \ref{deltaX}.

\subsubsection{Delta}

In case of unstable particles the pole of the dressed propagator
is located in the complex plane. We choose the renormalized mass of the
delta resonance as the pole position of its dressed propagator in the chiral limit,
that is we apply a generalization of the complex-mass scheme introduced originally
for the Standard Model \cite{Stuart:1990,Denner:1999gp}. Note that the non-trivial issue of
unitarity within the complex-mass scheme has been discussed in Ref.~\cite{Bauer:2012gn}
and studied in more details in Ref.~\cite{Denner:2014zga}.

The propagator of the Rarita-Schwinger field corresponding to the Lagrangian
of Eq.~(\ref{Dlagr0})  for $A=-1$ in $d$ space-time dimensions has the form
\bea
i S_{0,ij}^{\mu\nu}(p)&=&-\frac{i(\slashed{p}+m_\Delta)}{p^2-m_\Delta^2+i0^+}
\Biggl[g^{\mu\nu}-\frac{1}{d-1}\,\gamma^\mu\gamma^\nu \nonumber\\
&+& \frac{1}{(d-1)m_\Delta}(p^\mu\gamma^\nu-\gamma^\mu p^\nu)
-\frac{d-2}{(d-1)m_\Delta^2}p^\mu p^\nu\Biggr]\xi_{ij}^{\frac{3}{2}}\ .
\eea
Using notations of Ref.~\cite{Hacker:2005fh} and defining $ i \Sigma^{\mu\nu}$ as the sum of one-particle irreducible diagrams
contributing to the delta two-point function,
we parameterize the self-energy of the $\Delta$ as
\begin{eqnarray}
\Sigma^{\mu\nu} & = & \Sigma_1\, g^{\mu\nu}+\Sigma_2\,
\gamma^{\mu}\gamma^{\nu}+\Sigma_3\,p^{\mu}\gamma^{\nu}+\Sigma_4\,
\gamma^{\mu}p^{\nu}+\Sigma_5\,p^{\mu}p^{\nu} + \Sigma_6\, p\hspace{-.45em}/\hspace{.1em}
g^{\mu\nu}\nonumber\\
&+&\Sigma_7\,p\hspace{-.45em}/\hspace{.1em}
\gamma^{\mu}\gamma^{\nu}+\Sigma_8\,
p\hspace{-.45em}/\hspace{.1em} p^{\mu}\gamma^{\nu}+\Sigma_9\,
p\hspace{-.45em}/\hspace{.1em}
\gamma^{\mu}p^{\nu}+\Sigma_{10}\,p\hspace{-.45em}/\hspace{.1em}
p^\mu p^\nu,
\label{DseParametrization}
\end{eqnarray}
where the $\Sigma_i$ are functions of $p^2$.
The dressed delta propagator has the form
\begin{equation}
i\,S^{\mu\nu}_{ij}(p)=\frac{-i \,g^{\mu\nu} \xi_{ij}^{\frac{3}{2}} }
{\slashed p-m_\Delta -\Sigma_1^\Delta(p^2) -\slashed p \,\Sigma_6^\Delta(p^2)}+{\rm NP}
=\frac{-i \,g^{\mu\nu} \xi_{ij}^{\frac{3}{2}}\, \mathcal{Z}_\Delta}{\slashed p-z_\Delta}+{\rm NP},
\label{DressedDPr}
\end{equation}
where the pole position $z_\Delta$ of the $\Delta$-propagator is obtained by
solving the equation
\begin{equation}
z_\Delta - m_\Delta -\Sigma_1^\Delta(z_\Delta^2)-z_\Delta\, \Sigma_6^\Delta(z_\Delta^2)=0\,.
\label{poleequation}
\end{equation}
The leading tree-order contribution to the delta self-energy is shown in
diagram (d) in Fig.~\ref{fig:m} and the leading loop contributions are given by
diagrams in Figs.~\ref{fig:m} (e) and (f). Similarly to the nucleon case,
the delta wave function renormalization constant is obtained via
\bea
\mathcal{Z}_\Delta&=&\left[\Sigma_6^{\Delta,{\rm loop}}(s)
+2s\frac{\partial}{\partial s}\Sigma_6^{\Delta,{\rm loop}}(s)
+2\slashed{p}\frac{\partial}{\partial s}\Sigma_1^{\Delta,{\rm loop}}(s)
\right]_{\slashed{p}=z_\Delta}\ .
\eea
The explicit expressions of $\Sigma_1^{\Delta,{\rm loop}}$ and $\Sigma_6^{\Delta,{\rm loop}}$
are given in the Appendix~\ref{sec.SE}. Renormalization of  the one-loop delta mass is carried
out using Eq.~(\ref{eq.EOMSshifts}) with the counter terms shown in Appendix~\ref{deltaX}.

\subsection{Coupling constants of the leading order interactions}
\subsubsection{Pion decay constant $F_{\pi}$}
For practical convenience, one often needs to replace the quantities in the
chiral limit, $F$, $g_R$ and $h_R$  by the physical ones, $F_\pi$, $g_{\pi N}$ and
$g_{\pi N\Delta}$, respectively.  At one-loop order the pion decay constant $F_\pi$ is
given via \cite{Gasser:1983yg}
\bea\label{eq.fpi}
F_\pi=F\left[1+\ell_4\frac{M^2}{F^2}+\frac{1}{F^2}\mH_\pi\right] ,
\eea
which is renormalized in the standard way, i.e. $\ell_4= \ell_{4R}-{R}/({16\pi^2})$.

\subsubsection{Pion-nucleon coupling constant $g_{\pi N}$}

\begin{figure}[t]
\begin{center}
\epsfig{file=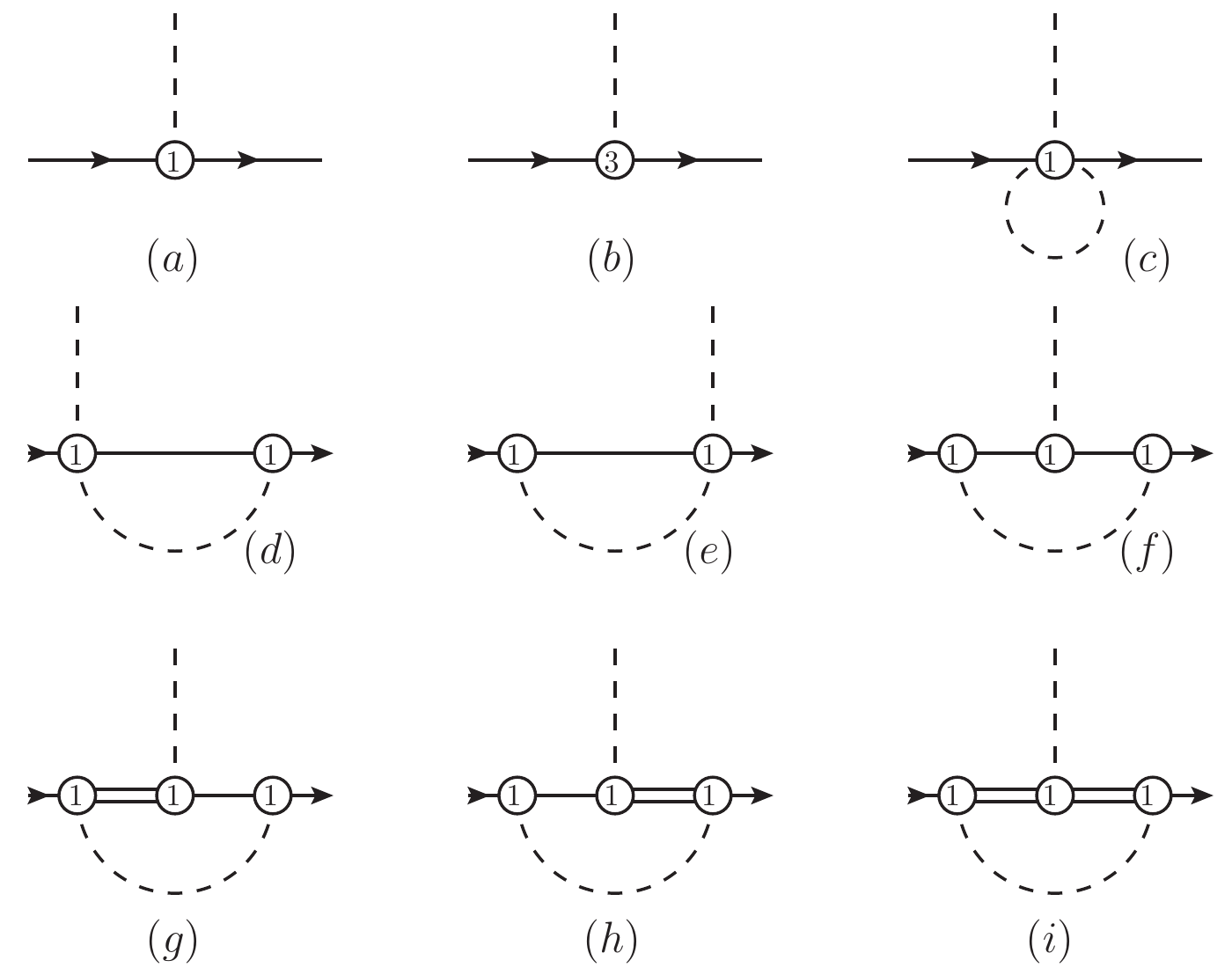,scale=0.6}
\caption{Tree and one-loop diagrams contributing to the $g_{\pi N}$
coupling constant up to  order $O(p^3)$.}
\label{fig:gPiN}
\end{center}
\end{figure}

In the isospin limit $m_u=m_d=\hat m$, where $m_u$ and $m_d$ are the masses of the $u$
and $d$ quarks, respectively, the matrix element of the
pseudoscalar density evaluated between one-nucleon states can be  parameterized as
\begin{equation}
\hat m \langle N(p')|P^{a}(0) |N(p)\rangle = \frac{M_\pi^2 F_\pi}{M_\pi^2-q^2}
G_{\pi N} (q^2) \,i\,\bar u(p') \gamma^5 \tau^a u(p)~,
\label{AFFpar}
\end{equation}
where $q=p'-p$. $G_{\pi N}(q^2)$ is called the pion-nucleon form factor and its
value at $q^2=M_\pi^2$ defines the pion-nucleon coupling constant $g_{\pi N}
=G_{\pi N}(M_\pi^2)$. Tree and one-loop diagrams up to order ${\cal O}(p^3)$
contributing to the $\pi N N$  vertex function are shown in  Fig.~\ref{fig:gPiN}.
The result up to leading one-loop order can be written as
\bea\label{eq.gpiNN}
g_{\pi N}=\frac{\ga\, m_N}{F_\pi}(1+g_{\pi N}^{(2)})\ ,\qquad
g_{\pi N}^{(2)}=\frac{2(d_{18}-2d_{16})M_\pi^2}{F_\pi^2}+g_{\pi N}^{(2),{\rm loop}}\ ,
\eea
where $g_{\pi N}^{(2),{\rm loop}}$ represents the very lengthy loop contribution which is
not given explicitly in this paper but obtainable from the authors (also the loop
contributions to $g_{\pi N\Delta}$ coupling, discussed below). The coupling
$g_{\pi N}$ is renormalized in the EOMS scheme and we have checked that the
divergences and PCBTs from the loop contributions can indeed be canceled by counter
terms generated by $\ga$ and $d_{18}-2d_{16}$.

\subsubsection{Pion-nucleon-delta coupling constant $g_{\pi N\Delta}$}

\begin{figure}[t]
\begin{center}
\epsfig{file=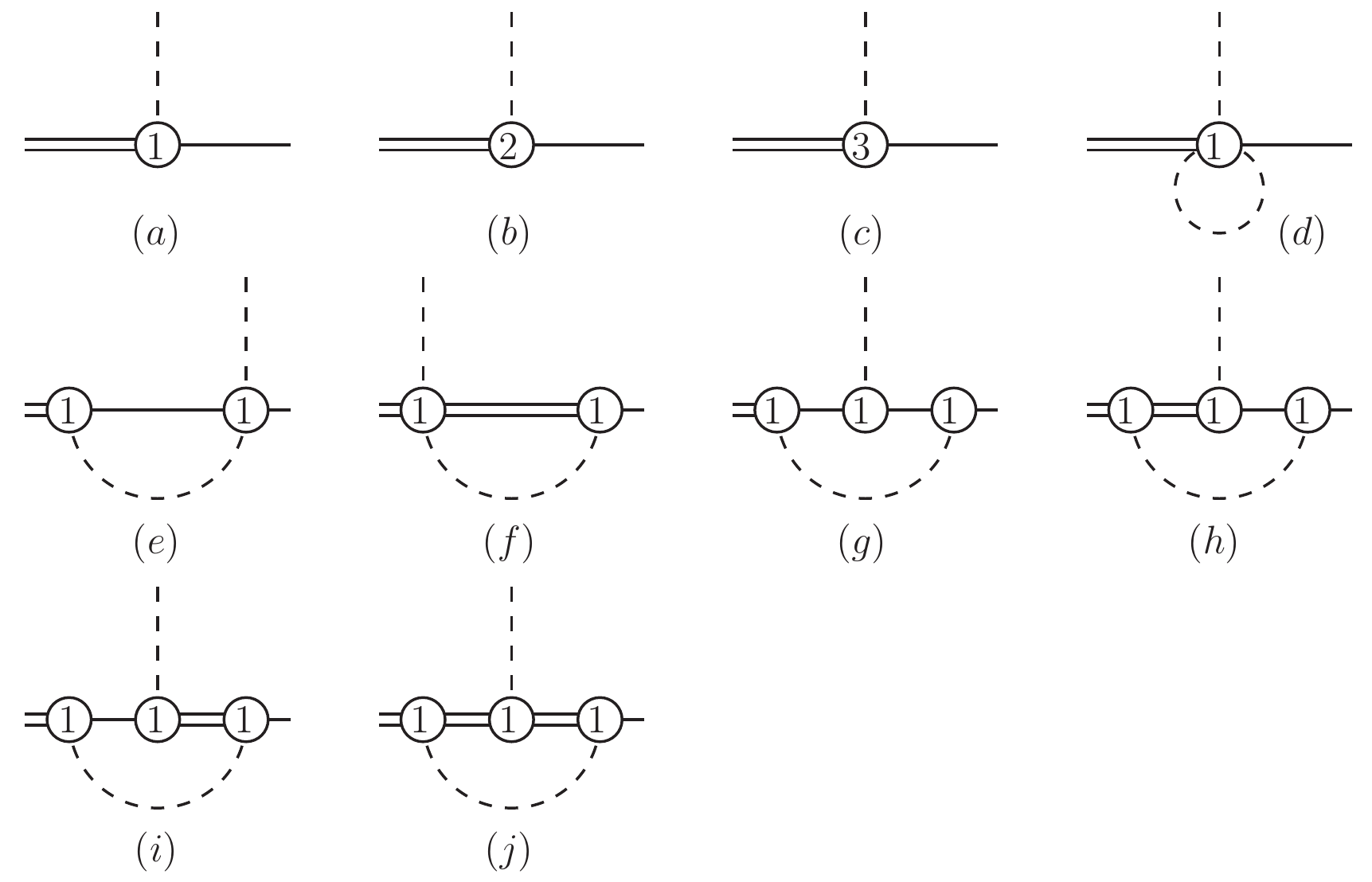,scale=0.6}
\caption{Tree and one-loop diagrams contributing to the $g_{\pi N\Delta}$ coupling
constant up to  order $O(p^3)$.}
\label{fig:gPiND}
\end{center}
\end{figure}

Following Ref.~\cite{Gegelia:2009py} we define the ``physical'' coupling constant
$g_{\pi N\Delta}$ by considering the pion-nucleon-delta vertex function $\Gamma$ on
the complex mass-shell of the delta. Tree and one-loop diagrams up to  order
${\cal O}(p^3)$  contributing in the $\pi N \Delta$  vertex function are shown in
Fig.~\ref{fig:gPiND}.  With $t=q^2=(p^\prime-p)^2$, the form factor $g_{\pi N \Delta}(t)$
is defined by~\cite{Ellis:1997kc}
\bea
\bar{u}(p^\prime)\Gamma^{(i,\mu),a}(p,p^\prime,q)u_\mu(p) =
g_{\pi N\Delta}(t)\xi^{ia}\bar{u}(p^\prime)q^\mu u_\mu(p)~,
\eea
where $u_\mu(p)$ and  $\bar{u}(p^\prime)$ are the Dirac spinors of the delta and the nucleon,
respectively.  We define the ${\pi N\Delta}$ coupling by taking the external pion
on the mass-shell, i.e. $g_{\pi N\Delta}=g_{\pi N\Delta}(M_\pi^2)$.
One can redefine $\ha$ using Eq.~(\ref{eq:redefinition.h}) and by that the tree
contributions from Diagrams (b) and (c) in Fig.~\ref{fig:gPiND} can be absorbed.
Hence, the final expression of  $g_{\pi N\Delta}$ can be written as
\bea\label{eq.gpiND}
g_{\pi N\Delta}=\frac{\ha}{F_\pi}(1+g_{\pi N\Delta}^{(2),{\rm loop}})\ ,
\eea
with $\ha$ the redefined parameter (as specified in Eq.~(\ref{eq:redefinition.h}))
and $g_{\pi N\Delta}^{(2),{\rm loop}}$  the loop contribution. Like the renormalization of
$g_{\pi N}$, UV-divergences and PCBTs are canceled by counter terms generated by
$\ha$. As pointed out e.g. in Ref.~\cite{Ellis:1997kc}, the coupling $g_{\pi N\Delta}$ is
a complex-valued quantity.

\subsection{The $\pi N$ scattering amplitude}
According to the LSZ reduction formula, the $\pi N$ scattering amplitude
$T_{\pi N}$ is related to the amputated Greens function $\hat{T}_{\pi N}$,
which has been calculated in the above section, via
\bea
T_{\pi N}=\mathcal{Z}_\pi \mathcal{Z}_N\bar{u}(p^\prime)\hat{T}_{\pi N}u(p)\ ,
\eea
where $\mathcal{Z}_\pi$ and $\mathcal{Z}_N$ are the wave function renormalization constants 
of the pion and the nucleon, respectively.

 UV divergences and  power counting violating contributions of loop diagrams
contributing to the amplitudes which have to be subtracted are calculated
by applying the procedure outlined in details in Refs.~\cite{Fuchs:2003qc,Schindler:2003xv}.  
We do not give the expressions of these subtraction terms due to their large size.
As mentioned above these subtraction terms are canceled by counter terms generated
by parameters of the effective Lagrangian,  see Eq.~(\ref{eq.EOMSshifts})
and~(\ref{eq.MSbarshifts}).  After taking into account the contributions of the
counter terms, we obtain a finite amplitude respecting the power counting and
possessing the correct analytic behavior.

 Note that while we give the explicit form of the counter terms for $c_1$ in
the appendix, the ones for $c_{2,3,4}$ are too lengthy to be shown here.
However, one can display them as
 \bea
c_i= c_{i R}&+&\frac{1}{16\pi^2F^2}\left\{\mathcal{C}_1^i\mHO_N
+\mathcal{C}_2^i\mHO_\Delta+\mathcal{C}_3^i\mHO_{\pi\Delta}(\mn^2)
+\mathcal{C}_4^i\mHO_{\pi N}(\md^2)+\mathcal{C}_5^i\mHO_{\pi NN}(\mn^2,0,\md^2)\right.\nonumber\\
&&\left.\hspace{1.5cm}+\mathcal{C}_6^i\mHO_{\pi N\Delta}(\md^2,0,\mn^2)
+\mathcal{C}_7^i\mHO_{\pi \Delta\Delta}(\mn^2,0,\md^2)\right\} ,
\eea
with $i=2,3,4$ and $\mathcal{C}_{j=1,\cdots,7}^{i}$ being the corresponding coefficients.
Since the integrals like $\mHO_{\pi N}(m_\Delta^2)$ are complex, the $c_{2,3,4}$ are
renormalized to complex quantities. Note that  $c_1$ remains real after
renormalization in view of Eq.~(\ref{eq.EOMSc1}).

 Finally,  for the sake of practical use, the quantities in the chiral limit
contributing to the $\pi N$ amplitude at one-loop order,
such as $M$, $m_R$, $m_{\Delta R}$, $F$, $g_R$ and $h_R$, can be substituted
by their corresponding physical quantities specified by Eqs.~(\ref{eq.mpi}),
(\ref{eq.mN}), (\ref{poleequation}), (\ref{eq.fpi}), (\ref{eq.gpiNN}) and
(\ref{eq.gpiND}) as this leads to  differences beyond the accuracy of the
current calculation.

\section{%Partial wave
Phase shifts and threshold parameters}
\label{PWPSs}
Based on the $\pi N$ scattering amplitudes specified in the above section, we
calculate the phase shifts and threshold parameters. In what follows, we first
determine the unknown LECs involved in the $\pi N$ scattering amplitudes by fitting
to the phase shifts of the $S$- and $P$-waves. Then we predict the $D$- and $F$-wave
phase shifts and the threshold parameters using the determined LECs.

\subsection{Fitting procedure}
For the partial wave analysis of the pion-nucleon scattering amplitudes results of
several groups are available: Karlsruhe \cite{Koch:1980ay,Koch:1985bn},
Matsinos~\cite{Matsinos:2006sw} and GWU~\cite{SAID}. Unfortunately, none of these
groups provide uncertainties of their results.
Therefore, we prefer to perform fits to the phase shifts generated by the recent
Roy-Steiner-equation analysis (RS) of the $\pi N$ scattering~\cite{Hoferichter:2015hva},
where both the central values and the errors of results are given by Schenk-like
or conformal parameterizations. Note that this analysis also includes the most
up-to-date experimental information on the pion-nucleon scattering lengths.
For fitting we extract  the RS phase shifts equidistantly from the threshold
$W_{\rm th}=1078$~MeV to $W=1318$~MeV with a  step-size of $0.8$~MeV.  Furthermore,
at each fixed energy point, the central value of the phase shift is generated
randomly with a normal distribution $N(\mu,\sigma)$, where the mean value
$\mu$ and the standard deviation $\sigma$ are specified by the results of the
RS equation analysis, and the corresponding error to the central value is
assigned to be $\sigma$. This procedure generates a set of simulated data for the
chosen energy configuration which is suitable for fitting.
In order to obtain stable values for the LECs one  should repeat such a procedure
as well as fitting for a large number of times, which will generate a large number of
(central) values for the LECs and from which the mean values and the statistical errors
of the LECs can be determined. In order
to achieve this it is enough in our case to repeat the fitting procedure 100 times.
Note that our results are Gaussian and thus in fact any procedure using 
different fitting approaches would lead to the same central values and error bars. 
Our interest here is to obtain a value of the $\chi^2$ which has the usual
interpretation, namely a good $\chi^2$ corresponds to
a value close to 1. For fits done directly using the RS equations a good
$\chi^2$ would be close to zero.

In the fitting procedure, two $S$-waves, $S_{31}$ and $S_{11}$ and four $P$-waves,
$P_{31}$, $P_{11}$, $P_{33}$, $P_{13}$, are taken into account. We use
Eq.~(\ref{eq:psgeneralized}) to extract the phase shifts for $P_{33}$ partial wave,
where the Delta resonance is located. For the other partial waves, as discussed in
Section~\ref{sec:pshift}, the difference due to various unitarization procedures
appears at higher orders, and we use Eq.~(\ref{psold}). This is especially
advantageous for the $P_{11}$ partial wave,  since there is a numerical problem
when using Eq.~(\ref{eq:psgeneralized}) because the real part of the partial wave
amplitude vanishes for some energy close to the threshold.

There are eleven LECs (or independent combinations of them) involved in the $\pi N$
amplitudes in total: $c_1$, $c_2$, $c_3$, $c_4$, $d_{1}+d_2$, $d_3$, $d_5$, $d_{14}-d_{15}$,
$g_{\pi N}$, $g_{\pi N\Delta}$ and $g_1$. We fix $g_{\pi N}$ coupling at the central value 
of $g_{\pi N}^2/(4\pi)=13.69\pm0.20$,  which was recently 
obtained through the Goldberger-Miyazawa-Oehme sum rule~\cite{Baru:2010xn,Baru:2011bw}.
The renormalized couplings $c_{i=1,2,3,4}$ and $g_{\pi N\Delta}$ are complex due to absorption
of complex subtraction terms of loop diagrams. Nevertheless, for convenience, we
use real-valued renormalized $c_i$'s in our fitting procedure and the corresponding
imaginary parts of the subtraction terms are retained in the loop contributions rather than
absorbed in the $c_i$. As for $g_{\pi N\Delta}$, we define
\bea
h_A= g_{\pi N\Delta} F_\pi\, ,
\eea
which is more often used in BChPT and the large-$N_c$ relation yields
\bea\label{LNChA}
h_A=(3g_A)/2\sqrt{2}\simeq 1.35\ ,\qquad {\rm with} \qquad g_A=1.27\ .
\eea
Note that our notation differs from the one used in Ref.~\cite{Pascalutsa:2006up}
by a factor of 2.
Obviously, $h_A$ is also a complex coupling in the calculation up to NNLO.
However, one can just choose the real part ${\rm Re}[h_A]$ as a fitting parameter,
and the corresponding imaginary part  can be obtained using
Eq.~(\ref{eq.gpiND}). In practice, the involved loop integrals can be calculated with 
all the masses being specified in the next paragraph and hence
the imaginary part of $h_A$  is given by (up to higher order corrections)  
\bea
{\rm Im}[h_A]=\frac{\left(1.51 F_\pi^2 g_{\pi N}^2-1.84 {\rm Re}[h_A]^2 \right)
{\rm Re}[h_A]}{160 F_\pi^2\pi^2}\ .
\eea
Thus we are left with ten unknown real fitting parameters.
It is worth noting that $h_A$ appearing in the loop contributions can be
substituted by ${\rm Re}[h_A]$ since the difference caused is of higher order.

For the masses of the particles and the pion decay constant the following values
are employed throughout our fitting procedure:
$M_\pi=139~{\rm MeV}$, $ m_N=939~{\rm MeV}$, $z_\Delta=(1210-i\,50)~{\rm MeV}$,
$F_\pi=92.2~{\rm MeV}$.
We take the dimensional regularization scale $\mu=m_N$.
Here, $z_\Delta$ has been identified as the pole position of the dressed delta
propagator with its value given by PDG~\cite{Agashe:2014kda}.
Note that one can use ${\rm Re}[z_\Delta]$ in the loop contributions instead of $z_\Delta$
since the difference caused by this approximation is of higher-order,
at least order $\mathcal{O}(p^5)$.  This substitution guarantees that all arguments
of the required loop integrals are real (no arguments with complex momenta) and,
therefore, this enables us to calculate all one-loop integrals using the
programs for numerical evaluation OneLoop~\cite{vanHameren:2010cp} and
LoopTools~\cite{Hahn:1998yk}.   The fits below were performed using the Fortran package Minuit\cite{James:1975dr}.

%\begin{figure}[htbp]
\begin{figure}[t]
\begin{center}
\epsfig{file=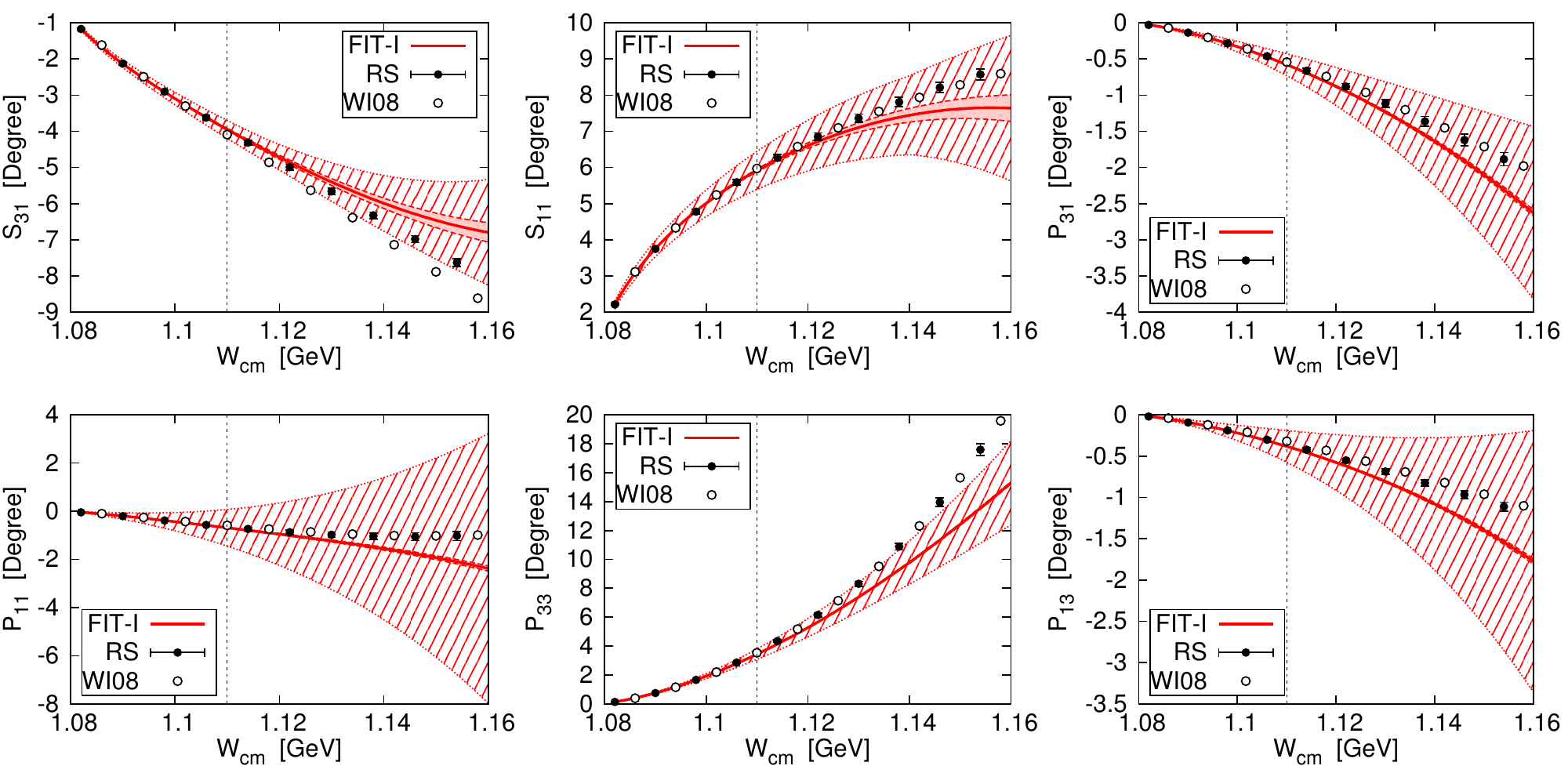,scale=0.75}
\caption{Phase shifts obtained from  delta-less BChPT by fitting RS
phase shifts in the c.m. energy range $[1082,1110]$~MeV (pion laboratory
momentum $q_\pi\in[36.1,108.4]$~MeV). Dots with error bars correspond to the
RS phase shifts, while circles without error bars stand for the GWU phase shifts.
The solid (red) lines represent the results of the current work. The red narrow error 
bands correspond to the uncertainties propagated from the errors of LECs using 
Eq.~(\ref{eq:errLECs}). The wide dashed error bands correspond to the theoretical 
uncertainties due to the truncation of the chiral series estimated by using 
Eq.~(\ref{eq:error}) proposed in Ref.~\cite{Epelbaum:2014efa}. }
\label{fig:RS01}
\end{center}
\end{figure}

\begin{figure}[t]
\begin{center}
\epsfig{file=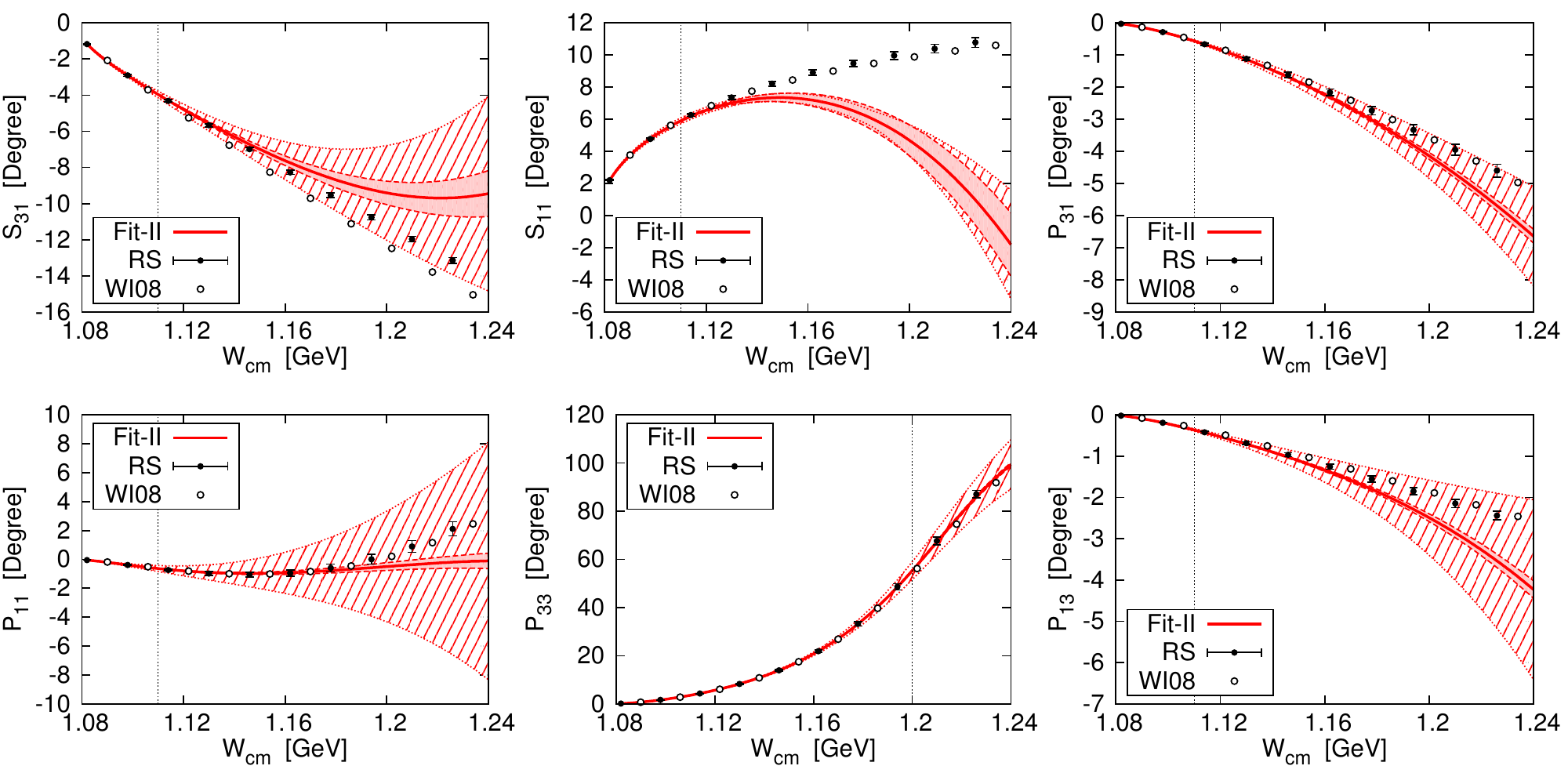,scale=0.75}
\caption{Phase shifts obtained from  BChPT with explicit delta degrees of freedom
in the tree diagrams corresponding to Fit-II. Dots with error bars stand for the RS 
phase shifts and circles without error bars represent the GWU phase shifts.
The solid (red) line represents the result of Fit II of the current work.
The red narrow error bands correspond to the uncertainties propagated from the errors of 
LECs using Eq.~(\ref{eq:errLECs}).
The wide dashed error bands correspond to the theoretical uncertainties due to the
truncation of the chiral series estimated by using Eq.~(\ref{eq:error})
proposed in Ref.~\cite{Epelbaum:2014efa}. 
}
\label{fig:RS02}
\end{center}
\end{figure}

\begin{figure}[t]
\begin{center}
\epsfig{file=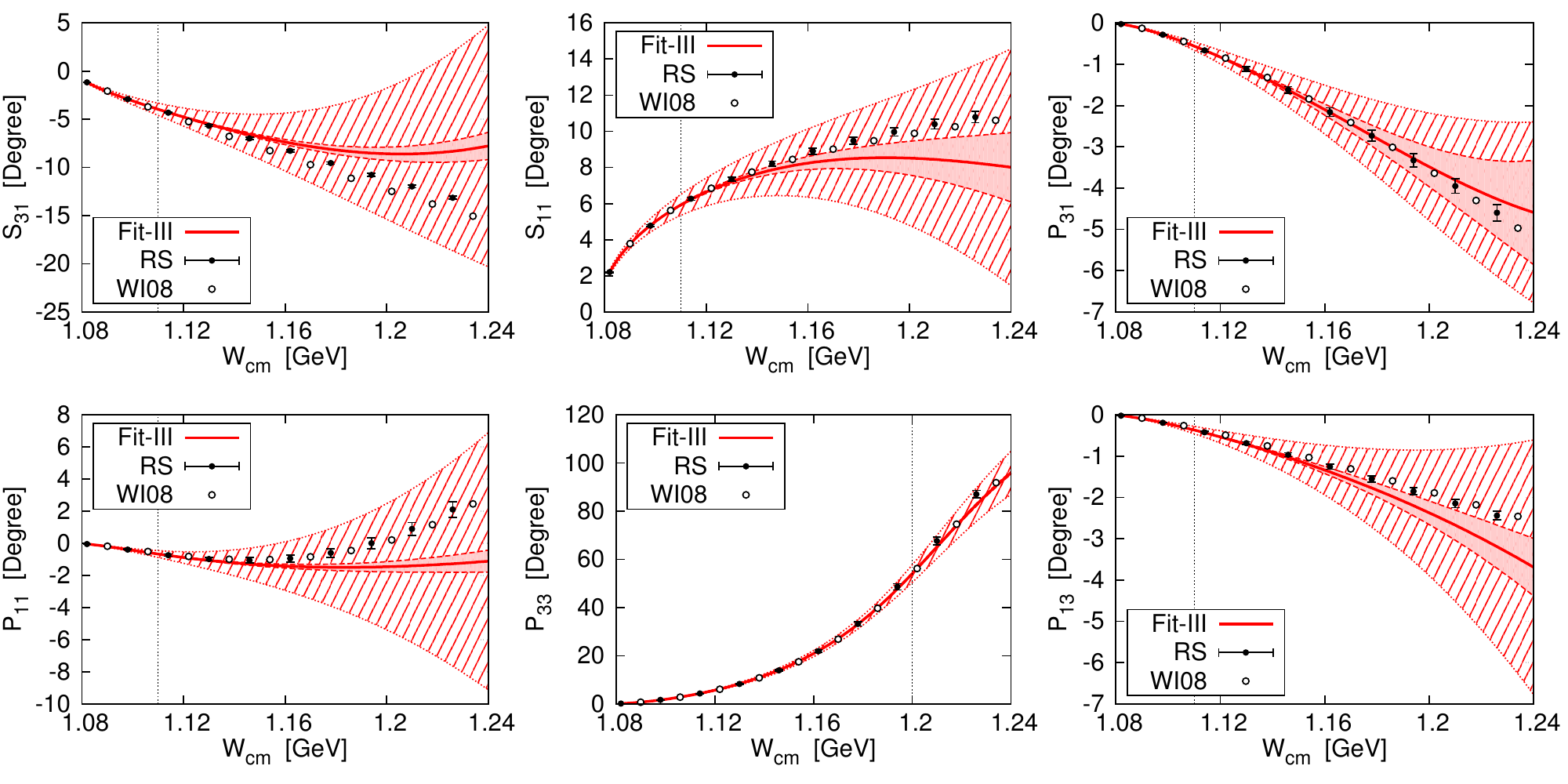,scale=0.75}
\caption{Phase shifts obtained from  BChPT with explicit delta degrees of freedom
corresponding to Fit-III. Dots with error bars stand for the RS phase shifts
and circles without error bars represent the GWU phase shifts. The solid (red)
line represents the result of Fit III of the current work.
The red narrow error bands correspond to the uncertainties propagated from the errors of 
LECs using Eq.~(\ref{eq:errLECs}).
The wide dashed error bands correspond to the theoretical uncertainties due to the
truncation of the chiral series estimated by using Eq.~(\ref{eq:error})
proposed in Ref.~\cite{Epelbaum:2014efa}. 
}
\label{fig:RS03}
\end{center}
\end{figure}

\subsection{Results}

\begin{table}[hbpt]
\caption{Values of the LECs for various fits to the RS phase shifts. The $c_i$
and $d_j$ are in units of GeV$^{-1}$ and GeV$^{-2}$, respectively.  The statistical and systematic uncertainties are shown in the first and the second brackets, respectively.}
\label{TabLECs}
\vspace{0.2cm}
\centering
\newcolumntype{C}{>{\centering\arraybackslash}X}
\newcolumntype{R}{>{\raggedleft\arraybackslash}X}
\renewcommand{\arraystretch}{1.4}
\begin{tabularx}{0.9\textwidth}{c|CCC}
\hline\hline
		&{Fit-I}			&{ Fit-II}				&{ Fit-III}			\\
LEC		&$N$ (i.e. $\slashed{\Delta}$)&$N$+LO $\Delta$& $N$+$\Delta$ \\
\hline
$c_{1}$		& $-1.22(2)(2)$		& $-0.99(2)(1)$		& $-1.31(2)(1)$		\\
$c_{2}$		& $3.58(3)(6)$		& $1.38(3)(1)$		& $0.78(4)(2)$		\\
$c_{3}$		& $-6.04(2)(9)$		& $-2.33(3)(1)$		& $-2.55(10)(7)$	\\
$c_{4}$		& $3.48(1)(3)$		& $1.71(2)(1)$		& $1.20(4)(2)$		\\
$d_{1+2}$		& $3.25(4)(9)$		& $0.14(4)(3)$		& $4.85(68)(64)$	\\
$d_{3}$		& $-2.88(8)(14)$	& $-0.97(8)(15)$	& $-0.62(10)(15)$	\\
$d_{5}$		& $-0.15(6)(14)$	& $0.39(6)(11)$		& $-0.93(11)(15)$	\\
$d_{14-15}$& $-6.19(7)(12)$		& $-1.08(8)(3)$		& $5.54(2.79)(2.01)$           \\
$g_{\pi N}$	& $13.12^\ast$		& $13.12^\ast$		& $13.12^\ast$	                    \\
$h_A$& $-$					& $1.28(1)(1)$		& $1.42(1)(1)-i\,0.16(1)(1)$   \\
$g_1$		& $-$			& $-$			& $-1.21(46)(39)$	\\
\hline
${{\chi}^2}/{\rm dof}$&$\frac{272.0(23.7)}{216-8}=1.31(11)$ &$\frac{339.8(27.4)}{328-9}=1.07(9)$&$\frac{373.8(29.9)}{328-10}=1.18(9)$\\
\hline\hline
\end{tabularx}
\end{table}

\begin{table}[hbpt]
\caption{Correlation and covariance coefficients for the fits. The upper and lower triangles 
correspond to the correlations (in unit of $10^{-2}$) and covariances (in unit of 
$10^{-4}$), respectively. }
\label{TabCor1}
%\vspace{0.2cm}
\centering
\newcolumntype{C}{>{\centering\arraybackslash}X}
\newcolumntype{R}{>{\raggedleft\arraybackslash}X}
\renewcommand{\arraystretch}{1.03}
\begin{tabularx}{\textwidth}{c|CCCCCCCCCC}
\hline\hline
\parbox[c]{2.0cm}{Fit-I}			&$c_1$	&$c_2$	&$c_3$	&$c_4$	& $d_{1+2}$ 	& $d_3$	&$d_5$ &$d_{14-15}$	\\
\hline
$c_{1}$		& $6$	& $83$	& $35$	 &$-10$ &$-14$		&$-18$	&$24$&$16$	\\
\cline{3-3}$c_{2}$&$7$ 		& $12$	& $-22$	 &$28$ 	&$28$		&$-29$	&$16$&$-20$	\\
\cline{4-4}$c_{3}$		& $2$ 		& $-2$		& $4$	 &$-62$	 &$-73$		&$19$	&$13$&$65$	\\
\cline{5-5}$c_{4}$		& $1$		& $1$		& $-2$		 &$2$	 &$26$		&$-8$	&$-4$&$-19$	\\
\cline{6-6}$d_{1+2}$	& $-1$		& $4$		& $-6$		 &$1$	 	&$14$			&$-4$	&$-37$&$-77$	\\
\cline{7-7}$d_{3}$		& $-3$		& $-8$		& $3$		 &$-1$		 &$-1$			&$65$		&$-91$&$27$	\\
\cline{8-8}$d_{5}$		& $3$		& $3$		& $2$		 &$1$		 &$-8$			&$-42$		&$32$	     &$6$	\\
\cline{9-9}$d_{14-15}$& $3$		& $-5$		& $10$		 &$-2$		 &$-21$			&$15$		&$3$	     &$51$	\\
%\hline\hline
\end{tabularx}
\begin{tabularx}{\textwidth}{c|CCCCCCCCC}
\hline\hline
\parbox[c]{2.cm}{Fit-II}				&$c_1$	&$c_2$	&$c_3$	&$c_4$	& $d_{1+2}$ 	& $d_3$	&$d_5$ 	&$d_{14-15}$	&${\rm Re} [h_A]$		\\
\hline
$c_{1}$				& $6$	&$80$	& $36$	& $-16$ 	&$-17$		& $-17$	&$25$	& $19$			& $17$				 \\
\cline{3-3}$c_{2}$		& $6$		& $12$	& $-26$	 &$31$ 	& $31$		& $-28$	& $14$	& $-25$			& $-27$			\\
\cline{4-4}$c_{3}$		& $2$		& $-2$		& $7$	 &$-72$	 &$-78$		& $18$	& $17$	&$74$			&$80$			\\
\cline{5-5}$c_{4}$		& $-1$		& $2$		& $-3$		 &$2$	 & $42$		&$-10$	&$-10$	&$-38$			& $-56$				\\
\cline{6-6}$d_{1+2}$	& $-2$		& $4$		& $-9$		 & $3$		 &$16$		& $-3$	&$-39$	&$-80$			& $-64$			\\
\cline{7-7}$d_{3}$		& $-3$		& $-8$		& $4$		& $-1$		 &$-1$			&$64$		&$-90$	&$25$			&$18$			\\
\cline{8-8}$d_{5}$		& $3$		& $3$		& $3$		 & $-1$		 & $-9$			& $-41$		&$33$	     	&$10$			&$11$			\\
\cline{9-9}$d_{14-15}$& $3$		& $-7$		& $16$		 & $-5$		 & $-25$			& $16$		&$4$		&$60$				&$63$			\\
\cline{10-10}${\rm Re} [h_A]$& $1$		& $1$		& $1$		 & $1$		& $-1$			& $1$		&$1$		&$2$				&$1$				\\
%\hline%\hline
\end{tabularx}
\begin{tabularx}{\textwidth}{c|CCCCCCCCCC}
\hline\hline
\parbox[c]{2.cm}{Fit-III}			&$c_1$	&$c_2$	&$c_3$	&$c_4$	& $d_{1+2}$ 	& $d_3$	&$d_5$ 	&$d_{14-15}$	&${\rm Re} [h_A]$	&$g_1$	\\
\hline
$c_{1}$		& $6$	& $67$	& $-7$	&$7$ 	&$18$		&$-8$	&$-4$	&$-16$			&$5$			& $-18$	 \\
\cline{3-3}$c_{2}$		& $7$		& $19$	& $-44$	 &$56$ 	&$27$		&$-20$	&$-7$	&$18$			&$-61$			&$-25$\\
\cline{4-4}$c_{3}$		& $-2$		& $-19$ 		& $91$	 &$-93$	 &$-86$		&$-37$	&$68$	&$-88$			&$51$			&$91$\\
\cline{5-5}$c_{4}$		& $1$		& $9$		& $-33$		 &$14$	 &$73$		&$26$	&$-55$	&$72$			&$-64$			&$-77$\\
\cline{6-6}$d_{1+2}$	& $30$		& $80$		& $-562$		 & $186$		 &$4650$		&$44$	&$-85$	&$87$			&$-11$			&$-97$\\
\cline{7-7}$d_{3}$		& $-2$		& $-8$		& $-34$		& $9$		 & $284$			&$90$		&$-79$	&$55$			&$15$			&$-50$\\
\cline{8-8}$d_{5}$		& $-1$		& $-3$		& $68$		 & $-22$		 & $-609$			& $-79$		&$110$	     	&$-73$			&$-3$			&$81$\\
\cline{9-9}$d_{14-15}$& $105$		& $213$		& $-2339$		 & $752$		 & $16597$		& $1456$		&$-2142$		&$77827$				&$-11$			&$-97$\\
\cline{10-10}${\rm Re} [h_A]$& $1$		& $-1$		& $2$		 & $-1$		& $-4$			& $1$		&$1$			&$-16$				&$1$				&$15$\\
\cline{11-11}$g_1$		  &  $-19$		& $-50$		 & $403$		 & $-133$		&  $-3053$		&  $-219$		 & $393$		&$-12474$		 	& $4$&$2142$	\\
\hline\hline
\end{tabularx}
\end{table}

The fitted LECs for three different cases are given in Table~\ref{TabLECs}, where
the statistical and systematic uncertainties are shown in the
first and second brackets behind the central values, respectively.
The systematic uncertainties represent
the effects of varying the fitting ranges, which will be discussed later on
in this section. The covariance and correlation matrices between the LECs are given in
Table~\ref{TabCor1}. They are calculated using the standard formulae 
\bea
{\rm Cor}(x_i,x_j)&=&\frac{{\rm Cov}(i,j)}{\sqrt{{\rm Cov}(x_i,x_i) {\rm Cov}(x_j,x_j)}}\ ,\nonumber\\
{\rm Cov}(x_i,x_j)&=&\frac{(\bf{x_i}-\bar{\bf{x_i}})^T\cdot (\bf{x_j}-\bar{\bf{x_j}})}{N-1}\nonumber\ ,
\eea
where ${\bf x_i}$ is the vector of $N$ central values of the LECs obtained from the 
fitting of our pseudo-data as explained before, for our case $N=100$. 
There are strong correlations between some LECs. Results of our fits are displayed and
compared with the phase shifts of the RS analysis as well as GWU analysis in
Figures~\ref{fig:RS01}, \ref{fig:RS02} and \ref{fig:RS03}.  The red narrow bands stand 
for the uncertainties propagated from the errors of the LECs using
\bea\label{eq:errLECs}
\delta\,\mathcal{O}_{\rm LEC}=\bigg\{\left[\frac{\partial\mathcal{O}(\bar{x}_i)}{\partial x_i}
\right]^2\left(\delta x_i\right)^2+{\rm Cor}(x_i,x_j)
\left[\frac{\partial\mathcal{O}(\bar{x}_i)}{\partial x_i}\right]
\left[\frac{\partial\mathcal{O}(\bar{x}_j)}{\partial x_j}\right]
\delta x_i\delta x_j\bigg\}^{\frac{1}{2}}\,, \ i\neq j\,,
\eea
where $\mathcal{O}$ is any observables under consideration and the summation over the 
repeated indices is meant.  Note that the contributions from statistical and systematic 
errors of the LECs to the error of  the observable $\mathcal{O}$ are added in quadrature.
The dashed wide bands represent
uncertainties estimated by truncation of the chiral series for the central values
of LECs, using the method which was proposed in Ref.~\cite{Epelbaum:2014efa}.
To be specific, the uncertainty $\delta\,\mathcal{O}^{(n)}$ of a prediction for
an observable $\mathcal{O}$ up to $O(p^n)$ is assigned to be
\bea\label{eq:error}
\delta\,\mathcal{O}^{(n)}_{\rm theo.}={\rm max}\big(| \mathcal{O}^{(n_{\rm LO})} | Q^{n-n_{\rm LO}+1},
\{|\mathcal{O}^{(k)}-\mathcal{O}^{(j)}|Q^{n-j}\}\big)\, ,\qquad n_{\rm LO}\leq j< k\leq n\, ,
\eea
with $Q=\omega_q/\Lambda_b$ where $\omega_q$ and $\Lambda_b$ are the pion energy
in the center-of-mass frame and the breakdown scale of the chiral expansion,
respectively.  For the delta-full case, we choose to employ $\Lambda_b\sim 0.6$~GeV following
Ref.~\cite{Epelbaum:2014efa}, which is lower than the scale of the chiral symmetry
breaking $\Lambda_\chi\sim 4\pi F_\pi\sim 1$~GeV. The lightest particle we do
not include explicitly is the Roper resonance $N^*(1440)$ and its mass differs from the nucleon mass
by about $0.5$~GeV. Our choice of $\Lambda_b$ is close to that number. Similarly, for 
the delta-less case, $\Lambda_b$ is taken $0.4$~GeV due to the nucleon-delta mass difference. 
For more discussions on the choice of $\Lambda_b$ see Ref.~\cite{Siemens:2016hdi}.
Now let us proceed with the details of the three different fits.

	Fit-I corresponds to  the delta-less case and is performed up to
$W_{\rm max}$=$1.11$~GeV.  This maximal energy is chosen according to the
following criterions: I) the average $\overline{\chi^2}$ per degree of
freedom ($\overline{\chi^2}/{\rm d.o.f}$) for the 100-times fits is around $1.0,$
 II) the average $\overline{\chi^2}$  increases rapidly if one takes larger
$W_{\rm max}$. 
For Fit-I, we get results similar to those
obtained by fitting to the phase shifts of partial wave analysis by GWU group up to 
$1.13$~GeV~\cite{Alarcon:2012kn,Chen:2012nx}. There exist slight differences
between our current results and the previous ones~\cite{Alarcon:2012kn,Chen:2012nx}
due to the fact that different data (RS data versus GWU data) were fitted and
the fitting ranges are not the same. Besides, they have one more fitting
parameter $d_{18}$, which is related to $g_{\pi N}$ by making use of the Goldberber-Treiman
relation at NNLO. Our plots for Fit-I are shown in Figure~\ref{fig:RS01}. The error
bands in $P_{33}$ and $S_{31}$ partial waves do not cover the RS and GWU data beyond
the fitting range, which suggests that $\Lambda_b=0.4$~GeV underestimates the 
theoretical errors for these partial waves.

	Adding the delta degree of freedom should mostly improve the
description of the $P_{33}$ wave in the $\Delta$-resonance region. We thus performed
two fits (Fit-II and Fit-III) using 1.2~GeV as $W_{\rm max}$  for the $P_{33}$ 
partial wave and 1.11~GeV for the other five partial waves - the same value as for Fit I. 
Fit II is done by adding the LO Born-term contribution of the 
delta-exchange diagrams to the delta-less case and serves only the purpose 
of estimating the effect of the loop diagrams. Our plots for Fit-II are shown
in Figure~\ref{fig:RS02}. Since only the tree order contributions of the delta
are included, $h_A$ is a real parameter and meanwhile $g_1$ does not show up
in Fit-II. Compared to the strategies in Refs.~\cite{Alarcon:2012kn,Chen:2012nx},
in the current work the complex pole position of the delta propagator rather than
the real mass is incorporated in a systematic way and hence the effect of the
delta width is included explicitly. We obtain better results with smaller
uncertainties than the previous studies, for instance, the large errors in
$d_{14}-d_{15}$ are substantially reduced.
	
	Fit-III is done (up to 1.2~GeV for $P_{33}$ and up to $1.11$~GeV for the other waves) 
with the full contributions of pions, nucleons
and deltas up to NNLO. The obtained LECs of Fit-III are different from those of Fit-II
due to the inclusion of contributions of loop diagrams involving delta lines. Note that
all the $c_i$ and most of the higher order LECs are of natural size.
Our plots for Fit-III are shown in Figure~\ref{fig:RS03}.  Compared to the
plots in Figure~\ref{fig:RS02}, although both fits describe well the phase
shifts in the fitting range, Fit-III improves the predictions beyond fitting ranges 
in most of the partial waves, especially for the $S_{11}$ wave. 
The larger theoretical error in Figure~\ref{fig:RS03} compared to Figure~\ref{fig:RS02} is 
due to the large contributions of delta-loop diagrams, which are not taken into account in 
estimation of the theoretical error of Fit-II using Eq.~(\ref{eq:error}).

As one can see from Table~\ref{TabLECs}, the imaginary part of $h_A$ from Fit-III is small
compared to the corresponding real part ${\rm Re}[h_A]$ and our determination
for ${\rm Re}[h_A]$ is close to the large-$N_c$ prediction~(\ref{LNChA}).
The obtained $g_1$ for Fit-III is nearly consistent (within the error bars) with the 
corresponding large-$N_C$ result, $|g_1|=9g_A/5\simeq 2.28$.  As noted in 
Ref.~\cite{Fettes:2000bb}, $g_1$ appears only in the loop contribution, hence a precise 
determination of its value is not to be expected.
	
\begin{figure}[t]
\begin{center}
\epsfig{file=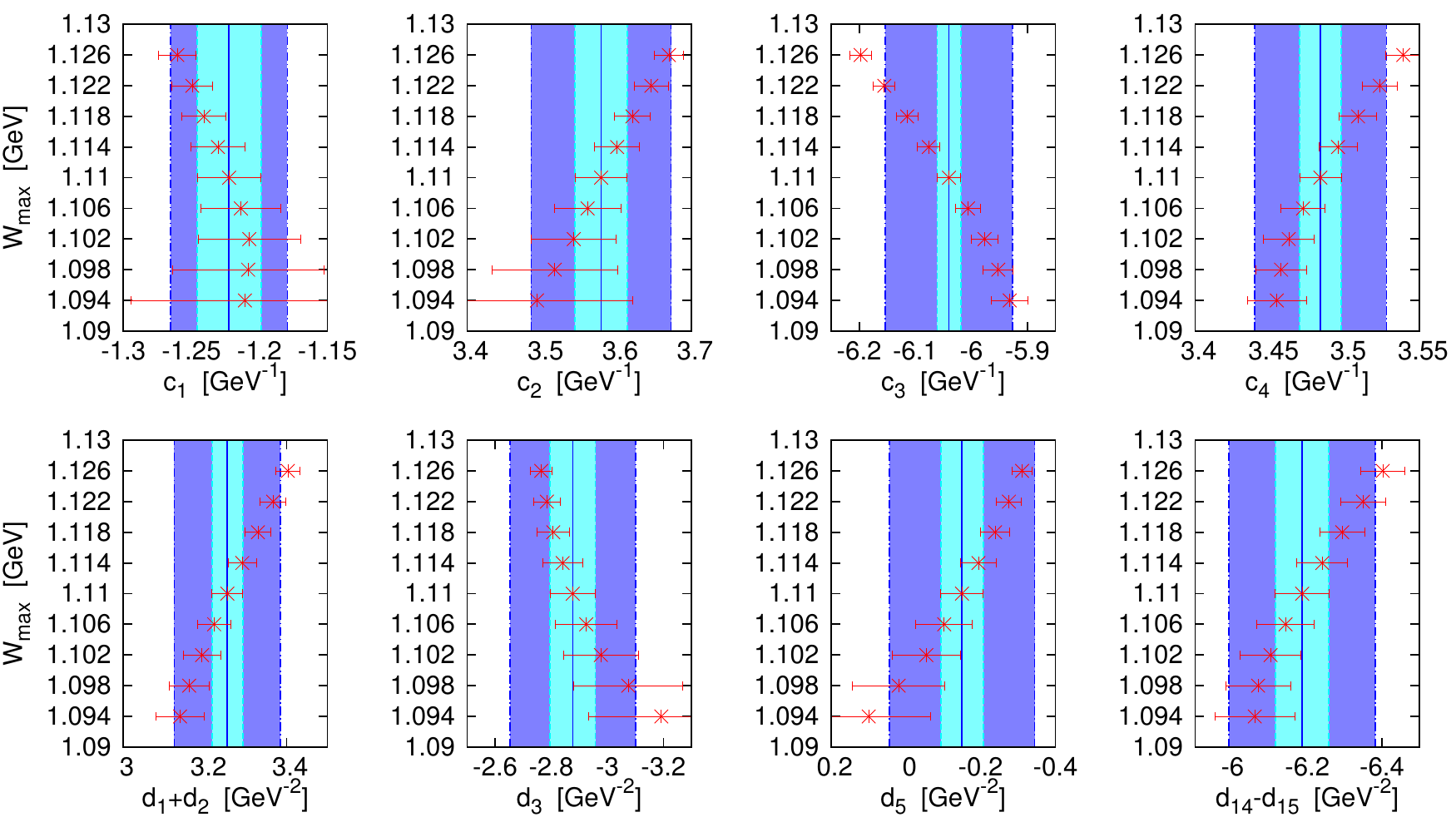,scale=0.8}
\caption{Demonstration of the effect of varying the fitting range on the fitted LECs.
The solid (blue) line with inner (cyan) band indicate the central value and the
statistical error which come from the preferred fit with $W_{\rm max}=1.11$~GeV.
The outer (blue) band is yielded by adding the systematic uncertainty which is
generated by the scatter of all results with different fitting ranges.
The error bars correspond to statistical errors of the fitting procedure.
}
\label{fig:syserror}
\end{center}
\end{figure}

	All the above three fits are done with their own preferred $W_{\rm max}$. However,
following Ref.~\cite{Durr:2014oba}, one can change those maximal energies around
the preferred $W_{\rm max}$ and redo the fits to see the influence on the obtained
LECs. For Fit-I, we made fits with  $W_{\rm max}=1.11\pm0.004$~GeV, $1.11\pm 0.008$~GeV, 
$1.11\pm 0.012$~GeV and $1.11\pm 0.016$~GeV in order to produce such kind of systematic 
errors to the LECs. For Fit-II and Fit-III, we keep the maximal energy at $1.2$~GeV for the
$P_{33}$ partial wave but vary it for the other waves as is the case for Fit-I.
A demonstration of how to obtain the systematic errors is given in Fig.~\ref{fig:syserror} for 
the case of Fit-I and analogous figures are obtained for other three fits. The obtained 
systematic errors to the LECs are shown in the second bracket in Table~\ref{TabLECs}. 

Note that all the presented fits have been done with the energy steps of $0.8$~MeV.
We have checked that the influence of varying the energy step
on the central values of the LECs is essentially negligible. Also the statistical errors decrease 
when the fitting range is extended keeping the energy step the same, see Fig.~\ref{fig:syserror}, 
or more fitting points are added in the same fitting range. However, we do not estimate such 
systematic uncertainties here.
	
\begin{figure}[t]
\begin{center}
\epsfig{file=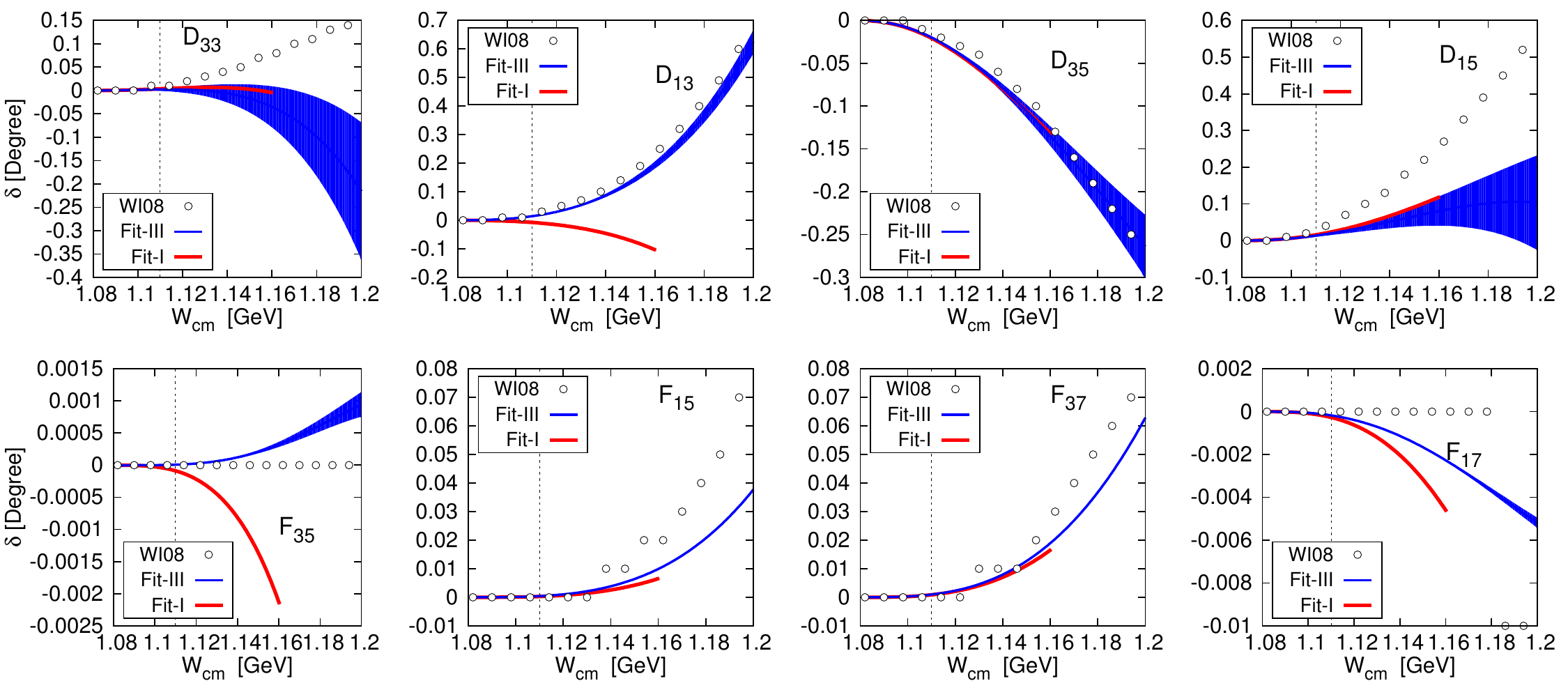,scale=0.7}
\caption{Phase shifts of the $D$ and $F$ partial waves obtained from the delta-less
and delta-full BChPT using the parameters of Fit-I (red line extending up to 1.16 GeV)
and Fit-III (blue line with band extending up to 1.2 GeV), respectively. 
The circles correspond
to phase shifts by the GWU group ~\cite{SAID}.
}
\label{fig:RS01DF}
\end{center}
\end{figure}

	Using the LECs obtained by fitting to the phase shifts of $S$- and $P$-waves, 
one can predict the phase shifts of  higher partial waves. In
Fig.~\ref{fig:RS01DF} we show the phase shifts of $D$ and $F$ partial
waves obtained using the parameters of Fit-I and Fit-III compared to the
results obtained by the GWU group ~\cite{SAID}. As expected, the predicted
phase shifts of higher partial waves are indeed small. Except for $D_{33}$
channel, our predictions agree qualitatively with the GWU results and
the  predictions of the delta-full theory are somewhat better than those of
the delta-less theory.

\begin{figure}[t]
\begin{center}
\epsfig{file=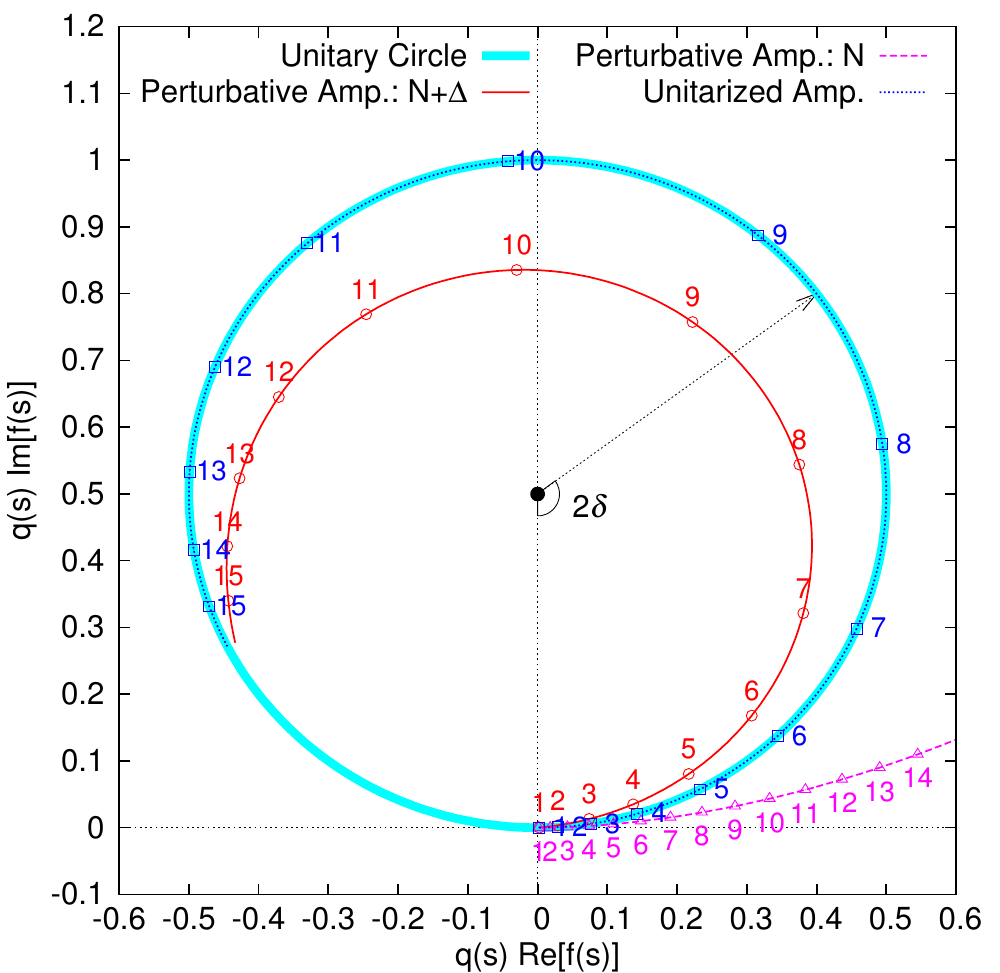,scale=0.7}
\caption{Argand diagram for the $P_{33}$ partial wave for the LECs of Fit-III.
For a detailed description see the text.
}
\label{fig:argands}
\end{center}
\end{figure}

Finally, in order to demonstrate how well Eq.~(\ref{eq:psgeneralized}) extracts phase
shifts from perturbative amplitudes, we draw the so-called Argand plot for the
$P_{33}$ partial wave in Fig.~\ref{fig:argands} by using the LECs of Fit-III.
In Fig.~\ref{fig:argands} the red solid and the magenta dashed lines
correspond, respectively, to  the full contribution ($\pi$+$N$+$\Delta$) and
the contribution  of the pion and nucleon ($\pi+N$) alone. As we can see,
the inclusion of the $\Delta$ contribution has a huge influence on improving the
unitarity constraints. The unitarized amplitude based on the full contribution is
represented by the blue dotted line, which is located on the unitary circle
(broad cyan solid circle), as expected. The energy corresponding to the
15th point is $W_{15}=1314$~MeV and the interval between two adjacent points
on the same line is $16$~MeV. The effect of Eq.~(\ref{eq:psgeneralized})
is to move the points of  the full NNLO perturbative amplitudes  to the
closest positions on the unitary circle.

\subsection{Scattering lengths and volumes}
At low energies, one can predict the threshold parameters based on the
above determined LECs.
For the partial wave with  angular momentum $\ell$ the general form of
the effective range expansion is given by
\bea
|{\bf p}|^{2\ell+1}{\rm cot}[\delta_{\ell\pm}^I]=\frac{1}{a_{\ell\pm}^I}
+\frac{1}{2}r_{\ell\pm}^I|{\bf p}|^2+\sum_{n=2}^\infty v_{n,\ell\pm}^I |{\bf p}|^{2n}\, ,
\label{ERE}
\eea
where ${\bf p}$ is the three-momentum of the nucleon in CMS frame, $a$ is the threshold parameter (e.g. scattering length for the $S$-wave
and scattering volume for the $P$-wave), $r$ is the effective range parameter,
and $v_n$ are the shape parameters. Using  Eq.~(\ref{ERE}) one can obtain the
threshold parameters as
\bea\label{eq.SL}
a_{\ell\pm}^I=\lim_{|{\bf p}|\to 0}\frac{\tan\delta_{\ell\pm}^I}{|{\bf p}|^{2\ell+1}}
=\lim_{|{\bf p}|\to 0}\frac{{\rm Re}f_{\ell\pm}^I(s)}{|{\bf p}|^{2\ell}}\, .
\eea
The second equality holds true due to the fact that the imaginary parts
of the partial wave amplitudes vanish faster at threshold. As
${{\rm Re}f_{\ell\pm}^I(s)}/{|{\bf p}|^{2\ell}}$ can not be computed numerically
exactly at the threshold we calculated its values for energies very close to
the threshold and then obtained the threshold parameters by extrapolating to
the threshold. Results of the threshold parameters corresponding to the three
different fits are presented in Table~\ref{Tab.thresholdpara}, together with the
determinations from the Roy-Steiner equation analysis~\cite{Hoferichter:2015hva} (and
the input from the analysis of pionic hydrogen and deuterium atom data).
The errors in the first brackets are propagated from the uncertainties of the
LECs, while the ones in the second brackets are obtained via Eq.~(\ref{eq:error}).
After taking the errors into consideration, all the obtained results agree well with
those of the Roy-Steiner equation analysis,  especially for Fit-III.

\begin{table}[hpbt]
\caption{Scattering lengths and volumes. The numbers in brackets correspond to the errors propagated from the uncertainties of LECs and the theoretical errors, respectively.}
\label{Tab.thresholdpara}
\vspace{0.2cm}
\centering
\newcolumntype{C}{>{\centering\arraybackslash}X}
\newcolumntype{R}{>{\raggedleft\arraybackslash}X}
\renewcommand{\arraystretch}{1.4}
\begin{tabularx}{0.9\textwidth}{c|CCCC}
\hline\hline
Threshold Para.				&Fit-I					&Fit-II					&Fit-III		  	 	& RS~\cite{Hoferichter:2015hva}\\
\hline
$a_{0+}^+~[10^{-3}M_\pi^{-1}]$ 	& $-0.6(7)(3.4)$			&$-1.1(7)(3.0)$				& $-0.5(7)(7.1)$	 		 &$-0.9(1.4)$\\
$a_{0+}^-~[10^{-3}M_\pi^{-1}]$		& $85.7(5)(3.3)$	 		&$85.8(4)(1.1)$				& $85.8(3)(1.0)$ 	 		&$85.4(9)$\\
$a_{1-}^+~[10^{-3}M_\pi^{-3}]$		& $-49.8(1.0)(15.9)$			&$-52.5(4)(4.7)$			& $-51.0(5)(6.7)$	 		&$-50.9(1.9)$\\
$a_{1-}^-~[10^{-3}M_\pi^{-3}]$		& $-9.7(3)(9.5)$			&$-11.3(3)(3.2)$			& $-9.5(2)(1.7)$	 		&$-9.9(1.2)$\\
$a_{1+}^+~[10^{-3}M_\pi^{-3}]$		& $139.9(1.8)(11.6)$			&$131.0(4)(4.0)$			& $131.5(5)(8.8)$	 		&$131.2(1.7)$\\
$a_{1+}^-~[10^{-3}M_\pi^{-3}]$		& $-84.0(6)(4.0)$			&$-80.3(1)(1.4)$			& $-80.4(2)(2.3)$	 		&$-80.3(1.1)$\\
\hline
\end{tabularx}
\end{table}

\section{Baryon sigma terms and the strangeness content of the nucleon\label{sec:sigmaterms}}
Sigma terms are interesting observables and important for understanding the sea-quark
structures of baryons. In particular, for the nucleon there are many studies of
the $\pi N$ $\sigma$-term, e.g., see
Refs.~\cite{Bernard:1993nj,Alarcon:2011zs,Hoferichter:2015dsa},
and of the strangeness content,
see Refs.~\cite{Gasser:1980sb,Borasoy:1996bx,Alarcon:2012nr,Ren:2014vea} for instance.
A high-precision determination of the $\sigma_{\pi N}$ was done from RS-equation
analysis based on the improved Cheng-Dashen low-energy theorem and
$\sigma_{\pi N}=(59.1\pm3.5)$~MeV was reported in Ref.~\cite{Hoferichter:2015dsa}.
In this section we discuss the $\sigma_{\pi N}$ based on our fitted results
obtained above. In order to estimate the strangeness content of the nucleon,
we perform the corresponding calculation in SU(3) BChPT. As byproducts,
the baryon sigma terms are also given.

\subsection{$\pi N$ sigma term}
The $\pi N$ sigma term $\sigma_{\pi N}$ can be obtained from the nucleon mass by
applying the Hellmann-Feynman theorem,
\bea\label{eq:SU2piN}
\sigma_{\pi N}=\hat{m}\frac{\partial m_N}{\partial \hat{m}},\quad
\hat{m}=\frac{(m_u+m_d)}{2}\ .
\eea
For practical convenience, using the expression of the nucleon mass $m_N$ given
in Section~\ref{sec:nucleon}, one can express $\sigma_{\pi N}$ as 
\bea
\sigma_{\pi N}=\underbrace{-77.28\, c_1}_{\rm LO}+\underbrace{(-11.72)\, 
g_A^2+(-6.55)\, {\rm Re}[h_A]^2}_{\rm NLO}\ ,
\eea
where the involved loop integrals have been computed numerically. Notice that the last term 
in this expression does not contribute in the calculation of the sigma terms of Fit-I and Fit-II.

The results for the pion-nucleon sigma term $\sigma_{\pi N}$
based on the different fitting results in the above section are shown in Table~\ref{tab:ST}, 
where contributions from different orders are also shown. The error in the first bracket
is propagated from the uncertainties of the fitted LECs using Eq.~(\ref{eq:errLECs}).
The error in the second
bracket is theoretical uncertainty estimated using Eq.~(\ref{eq:error}) and we employ 
$Q=M_\pi/\Lambda_b$ with $\Lambda_b=400$~MeV for the delta-less case and with $\Lambda_b=600$~MeV 
for the delta-full case. Note that here for Fit-II the theoretical error originating from the 
delta-loop contribution is also taken into account.
For easy comparison, the recent determination of the pion-nucleon sigma
term from the Roy-Steiner equations~\cite{Hoferichter:2015dsa} is also given. 

Our prediction for Fit-I, $\sigma_{\pi N}=74.8(2.2)(11.4)$~MeV, is marginally consistent with 
the RS determination when the large uncertainties are taken into account. 
For the delta-full case with LO delta contribution, applying the same unitarization as 
in Ref.~\cite{Alarcon:2011zs}, we obtain $\sigma_{\pi N}=60.1(1.6)(6.2)$~MeV based on Fit-II.
On the other hand, by including the explicit
delta width rather than generating it by unitarization as in
Ref.~\cite{Alarcon:2011zs}, we obtained $\sigma_{\pi N}=57.1(1.9)(7.0)$~MeV based
on Fit-II, which
appears to agree with the RS determination very well. As for Fit-III, our prediction 
$\sigma_{\pi N}=68.5(1.9)(7.6)$~MeV improves the delta-less result and within the error it overlaps
the value of the RS analysis. The large estimated theoretical error comes from the 
delta-loop diagram contributions by using Eq.~(\ref{eq:error}).

\begin{table}[hpbt]
\caption{The pion-nucleon sigma term in units of MeV.   The numbers in brackets correspond to the errors propagated from the uncertainties of LECs and the theoretical errors, respectively.}
\label{tab:ST}
\centering
\newcolumntype{C}{>{\centering\arraybackslash}X}
\newcolumntype{R}{>{\raggedleft\arraybackslash}X}
\renewcommand{\arraystretch}{1.3}
\begin{tabularx}{0.9\textwidth}{c|CCCC}
\hline\hline
		&{Fit-I}			&{Fit-II}			&{Fit-III}		& RS~\cite{Hoferichter:2015dsa}\\
%	&&&&\cite{Hoferichter:2015dsa}&\cite{Baru:2010xn,Baru:2011bw}\\
\hline
LO		&$94.3$				&$76.5$				&$101.2$			 &$-$\\
NLO		&$-19.5$				&$-19.5$				&$-32.7$			 &$-$\\
Sum		&$74.8(2.2)(11.4)$	&$57.1(1.9)(7.0)$	&$68.5(1.9)(7.6)$	 &$59.1(3.5)$\\
\hline\hline
\end{tabularx}
\end{table}

\subsection{The strangeness content of the nucleon and the sigma terms of baryons
from SU(3) BChPT}

Similarly to $\sigma_{\pi N}$ one can obtain the nucleon expectation value of
the operator $m_s \bar s s$ using the Hellmann-Feynman theorem,
\bea\label{eq:SU3sN}
\sigma_{sN}=m_s\frac{\partial m_N}{\partial m_s}\ ,
\eea
where $m_s$ is the mass of the strange quark. However, the nucleon mass  $m_N$
in the above equation should be calculated in SU(3) BChPT. We calculated the
masses of the baryon octet in SU(3) BChPT also including the baryon decuplet
as explicit degrees of freedom up to third order.  Details of this calculation
are specified in the Appendix~\ref{OSelfEnergy}. The other baryon sigma terms
are also obtainable by using formulas similar to Eqs.~(\ref{eq:SU2piN})
and ~(\ref{eq:SU3sN})  but with the nucleon mass being replaced by
$m_\Sigma$, $m_\Lambda$ or $m_\Xi$.

The strangeness content of the nucleon $y$ is defined through
\bea
y=\frac{2\hat{m}}{m_s}\frac{\sigma_{sN}}{\sigma_{\pi N}},
\eea
and the nucleon expectation value of the operator $\bar{u}u+\bar{d}d-2\bar{s}s$
(see e.g. \cite{Borasoy:1996bx}) is given by the following equation
\bea
\sigma_0={\sigma_{\pi N}} {(1-y)}.
\eea

In order to determine the strangeness content of the nucleon and the sigma terms
specified above, we first need to pin down the unknown LECs involved in the SU(3)
calculation. Therefore, we fit to the experimental values for $m_N$, $m_\Sigma$,
$m_\Lambda$ and $m_\Xi$ (taken from PDG~\cite{Agashe:2014kda})
as well as the RS determination of $\sigma_{\pi N}$, as given in the last column of
Table~\ref{tab:ST}.\footnote{Note that the experimental error, as specified in
PDG, for $m_\Lambda$ is extremely small and we assign an error of $0.1$~MeV to
$m_\Lambda$ in our fitting program in order to balance the $\chi^2$ contribution
with those originating from the other masses.} There are the following five
unknown LECs: the octet mass in the chiral limit $M_0$, the LECs corresponding to the
NLO mass splitting operators $b_0$, $b_D$ and $b_F$, and the LO
Goldstone-boson-octet-decuplet coupling constant $\mathscr{C}$. For the
other parameters we use the following values
\bea
D&=& 0.80 , \quad F=0.46 , \quad
M_\pi=0.139~{\rm GeV}, \quad M_K=0.494~{\rm GeV}, \nonumber\\
M_B&=&1.151~{\rm GeV}, \quad M_D=1.382~{\rm GeV},
\eea
with $M_B$ and $M_D$ being the averages of the physical masses of the octet and decuplet,
respectively. The mass of the $\eta$ meson is obtained from the Gell-Mann-Okubo
relation: $3M_\eta^2=4M_K^2-M_\pi^2$. Furthermore, $F_\phi=1.17 F_\pi$ with
$F_\pi=92.2$~MeV is used.

We performed two fits:
\begin{itemize}
\item Fit A: The octet baryon mass in the loops is fixed at $M_B$, and the mass
of the decuplet baryons to $m_D$.

\item Fit B: The octet baryon mass in the loops is set as the chiral limit
mass $M_0$, and the mass of the decuplet baryons to $m_D$.
\end{itemize}
If the chiral series of the baryon masses and the sigma terms converges well,
these two fits should differ slightly since the differences are of high order.
However, we obtain results with sizable differences (see $y$ and $\sigma_0$
together with the fitted parameters in Table~\ref{tab:LECsmas}) which implies
that the higher-order contributions are large.
Only when the theoretical errors, which is due to the truncation of the chiral
series using Eq.~(\ref{eq:error})  with $Q=M_\pi/\Lambda_\chi$, are taken into account,
these fit results overlap. The previous determinations are as follows:  $y=0.15(10)$ 
and $\sigma_0=33(5)$ of the NLO calculation~\cite{Gasser:1980sb}, $y=0.21(20)$ and $\sigma_0=36(7)$ 
of the NNLO calculation within HBChPT~\cite{Borasoy:1996bx}, $y=0.02(23)$ and $\sigma_0=58(8)$  
of the NNLO calculation within Covariant BChPT~\cite{Alarcon:2012nr}.
We therefore conclude that to this order in the chiral expansion, one is not able
to make a precise statement about the strangeness content of the nucleon.
Likewise, various values of $y$ calculated either directly in Lattice QCD, or indirectly 
using the octet baryon masses and sigma
terms obtained in Lattice QCD, are very different. Therefore we do not compare to those results.

We also predict the octet baryon masses and sigma terms in Table~\ref{tab:Mass}
and Table~\ref{tab:STs}, respectively. The errors for the masses and the ones
in the first brackets for the sigma terms are propagated from the uncertainties
of the LECs. For the sigma terms we also estimated the theoretical errors,
which are shown in the second brackets in Table~\ref{tab:STs}. As we can see,
the theoretical errors for $\sigma_{sN}$ are very large due to the bad convergence
of the chiral series in SU(3) BChPT. Note also that determinations of  
$\sigma_{sN}$ by different Lattice QCD collaborations vary in a large range, 
see e.g. Refs~\cite{Durr:2011mp,Yang:2015uis,Durr:2015dna,Horsley:2011wr,Freeman:2012ry,Junnarkar:2013ac,Oksuzian:2012rzb,Engelhardt:2012gd,Gong:2013vja,Abdel-Rehim:2016won}.

\begin{table}[hpbt]
\caption{LECs and strangeness content of the nucleon. Results
are obtained with $z=-1$ and $\mu=1$~GeV.  The numbers in brackets for $y$ and $\sigma_0$ correspond to the errors propagated from the uncertainties of LECs and the theoretical errors, respectively.}
\label{tab:LECsmas}
%\vspace{0.2cm}
\centering
\newcolumntype{C}{>{\centering\arraybackslash}X}
\newcolumntype{R}{>{\raggedleft\arraybackslash}X}
\renewcommand{\arraystretch}{1.4}
\begin{tabularx}{\textwidth}{c|ccccc|CC}
\hline\hline
&$M_0~({\rm GeV})$&$b_0$&$b_D$&$b_F$&$\mathscr{C}$&$y~~[\%]$&$\sigma_0~~[{\rm MeV}]$\\
\hline
Fit-A&$0.654(1)$	&$-1.155(1)$&$0.122(1)$	&$-0.359(1)$	&$1.495(5)$	&$1.2(6)(12.1)$&$58.4(4)(9.0)$\\
Fit-B&$0.622(2)$&$-0.956(1)$&$0.082(1)$&$-0.368(1)$&$2.289(20)$&$20.2(6)(6.5)$&$47.2(4)(5.2)$\\
\hline
\hline
\end{tabularx}
\end{table}

\begin{table}[hpbt]
\caption{Masses of octet baryons obtained with the LECs given in Table~\ref{tab:LECsmas}. The numbers in brackets correspond to the errors propagated from the uncertainties of LECs and the theoretical errors, respectively.}
\label{tab:Mass}
%\vspace{0.2cm}
\centering
\newcolumntype{C}{>{\centering\arraybackslash}X}
\newcolumntype{R}{>{\raggedleft\arraybackslash}X}
\renewcommand{\arraystretch}{1.4}
\begin{tabularx}{\textwidth}{c|CCCC}
\hline\hline
&$m_{ N}$	&$m_{\Lambda}$&$m_{\Sigma}$	&$m_{\Xi}$		\\
\hline
{Fit-A}	&$939.2(5.0)(61.8)$	&$1115.7(4.5)(77.1)$&$1186.0(4.5)(87.7)$&$1327.4(4.3)(97.6)$\\
{Fit-B} & $939.2(6.1)(33.7)$&$1115.7(5.6)(51.7)$&$1186.0(5.5)(55.6)$&$1327.4(5.4)(71.7)$\\
expt.&$938.925(645)$&$1115.683(6)$&$1193.15(4.30)$&$1318.28(3.43)$\\
\hline
\hline
\end{tabularx}
\end{table}

\begin{table}[hpbt]
\caption{Sigma terms obtained with the LECs given in Table~\ref{tab:LECsmas}. The numbers in brackets correspond to the errors propagated from the uncertainties of LECs and the theoretical errors, respectively.}
\label{tab:STs}
%\vspace{0.2cm}
\centering
\newcolumntype{C}{>{\centering\arraybackslash}X}
\newcolumntype{R}{>{\raggedleft\arraybackslash}X}
\renewcommand{\arraystretch}{1.4}
\begin{tabularx}{0.5\textwidth}{c|CC}
\hline\hline
&{Fit-A}	&{Fit-B}\\
\hline
$\sigma_{\pi N}$		&$59.1(2)(5.5)$	&$59.1(2)(3.6)$\\
$\sigma_{\pi\Lambda}$	&$46.9(2)(5.5)$	&$45.8(2)(3.6)$\\
$\sigma_{\pi\Sigma}$	&$38.6(2)(5.7)$	&$40.7(2)(3.7)$\\
$\sigma_{\pi\Xi}$		&$30.5(2)(5.6)$	&$30.0(2)(3.7)$\\
$\sigma_{s N}$			&$8.5(4.4)(86.6)$	&$144.7(4.6)(45.9)$\\
$\sigma_{s\Lambda}$	&$166.0(3.7)(106.3)$&$297.2(3.8)(69.1)$\\
$\sigma_{s\Sigma}$		&$203.6(3.9)(122.3)$&$355.4(4.0)(75.2)$\\
$\sigma_{s\Xi}$			&$342.5(3.4)(133.9)$&$479.0(3.4)(95.3)$\\
\hline
\hline
\end{tabularx}
\end{table}

\section{Summary and conclusion}
\label{summary}

	In this paper we presented the ${\cal O}(p^3)$ order calculation of
the pion-nucleon scattering amplitudes in the framework of BChPT including pions,
nucleons and deltas as explicit degrees of freedom. There are tree and one-loop
diagrams contributing at this order. We applied the EOMS renormalization
scheme to loop diagrams involving pion and nucleon lines only. For diagrams
with the delta lines in loops we used the complex-mass scheme which is
a generalization of the on-mass-shell scheme for unstable particles. That is
we subtracted the divergent pieces and the power counting violating contributions
of the loop diagrams by canceling them by counter terms generated by
splitting the bare parameters of the effective Lagrangian in renormalized couplings
and counter terms.

	We fitted the renormalized coupling constants to the $S$- and $P$-wave
phase shifts, which are randomly generated by using the results of the Roy-Steiner
equation analysis of Ref.~\cite{Hoferichter:2015hva} and hence are normally distributed
simulations. Both the phase shifts extracted from the RS equation analysis
and the GWU group analysis~\cite{SAID} are well described up to 1.11~GeV for
the delta-less case. For the delta-full case, the $P_{33}$ partial wave is fitted up 
to 1.20~GeV while the other partial waves up to 1.11~GeV. 
	
	Based on the obtained LECs, we predicted the $D$- and $F$-wave phase
shifts and compared them with the results given by the GWU group. We found that
our prediction for $D_{33}$ wave differs from the determination of the GWU group
while the predictions for other $D$ and $F$ waves agree well. Considering
the Argand plot for the $P_{33}$ partial wave we checked that the unitarized amplitude,
from which we extracted the phase shifts, is a good approximation to the amplitude
obtained by our perturbative calculation. At low energies, we  extracted the
threshold parameters and compared to the corresponding results of the Roy-Steiner
equation analysis obtaining satisfactory agreement.

In addition, we calculated the pion-nucleon sigma term. Our extractions of
$\sigma_{\pi N}$ based on the fitted LECs of Fit-I and Fit-III are 
consistent with the result of RS analysis $\sigma_{\pi N}=(59.1\pm3.5)$~MeV taking  
into account  the large errors of our determination.

In the end,  we also studied the strangeness content of the nucleon $y$ in SU(3)
BChPT. We first fixed all the involved SU(3) LECs by fitting to the experimental
octet baryon masses as well as the RS result
$\sigma_{\pi N}=(59.1\pm3.5)$~MeV~\cite{Hoferichter:2015dsa}. Two different
strategies were used. In principle, they should only slightly differ from
each other since the differences are due to  higher-order contributions.
However, because of the bad convergence properties of the SU(3) BChPT for these
quantities, we obtained two sets of predictions with  rather large discrepancies 
from each other.  Nevertheless, when the large uncertainties are taken into account, 
they are consistent with each other. Hence at the order one is working here, we unfortunately 
cannot disentangle a small from a large value of $y$.
Similar picture appears when one takes an overall view on the previous determinations 
from BChPT~\cite{Gasser:1980sb,Borasoy:1996bx,Alarcon:2012nr} and Lattice QCD~\cite{Durr:2011mp,Yang:2015uis,Durr:2015dna,Horsley:2011wr,Freeman:2012ry,Junnarkar:2013ac,Oksuzian:2012rzb,Engelhardt:2012gd,Gong:2013vja,Abdel-Rehim:2016won,Bali:2016lvx}.
Within these two strategies, we predict all
the octet baryon sigma terms as well.

\acknowledgments
We would like to thank Dalibor Djukanovic, Sebastien Descotes-Genon and Bastian Kubis 
for helpful discussions.
This work was supported in part by Georgian Shota Rustaveli National
Science Foundation (grant FR/417/6-100/14)and by the DFG (TR~16 and CRC~110).
The work of UGM was also supported by the Chinese Academy of Sciences (CAS) President’s
International Fellowship Initiative (PIFI) (Grant No. 2015VMA076).

\appendix
\section{Tree amplitudes of delta-exchange\label{app:treedelta}}
In what follows, the amplitudes corresponding to tree-order delta-exchange 
diagrams are given explicitly here.
\begin{itemize}
\item{Born-terms of} $\mathcal{O}(p^1)$ diagram (g):
\bea
A^+_{g}(s,t)&=&\frac{h^2}{9F^2\clm_\Delta^3(\clm_\Delta^2-s)}{\mathcal{F}_A(s,t)} ,\qquad A^-_g(s,t)=-\frac{1}{2}A^+_g(s,t) ,\nonumber\\
B^+_{g}(s,t)&=&-\frac{h^2}{9F^2\clm_\Delta^3(\clm_\Delta^2-s)}{\mathcal{F}_B(s,t)} ,\qquad B^-_g(s,t)=-\frac{1}{2}B^+_g(s,t) .
\eea
\item{Born-terms } of $\mathcal{O}(p^2)$ diagrams (h+i):
\bea
A^+_{hi}(s,t)&=&\frac{2h}{9F^2\clm_\Delta^3(\clm_\Delta^2-s)}\left\{ b_3\,\mathcal{G}_A(s,t)+b_8\frac{(s-m_N^2-M_\pi^2)}{2m_N}\mathcal{F}_A(s,t)\right\} ,\nonumber\\
B^+_{hi}(s,t)&=&-\frac{2h}{9F^2\clm_\Delta^3(\clm_\Delta^2-s)}\left\{ b_3\,\mathcal{G}_B(s,t)+b_8\frac{(s-m_N^2-M_\pi^2)}{2m_N}\mathcal{F}_B(s,t)\right\} ,
\nonumber\\
 A^-_{hi}(s,t)&=&-\frac{1}{2}A^+_{hi}(s,t) , \qquad B^-_{hi}(s,t)=-\frac{1}{2}B^+_{hi}(s,t).
\eea
\item{Born-terms of $\mathcal{O}(p^3)$} diagrams (j+k):
\bea
A^+_{jk}(s,t)&=&\frac{2h}{9F^2\clm_\Delta^3(\clm_\Delta^2-s)}\left\{-f_1\frac{s-m_N^2-M_\pi^2}{2m_N}\mathcal{G}_A(s,t)\right.\nonumber\\
&&\hspace{1cm}\left.+\left[-f_2\frac{(s-m_N^2-M_\pi)^2}{4m_N^2}+2(2f_4-f_5)M_\pi^2\right]\mathcal{F}_A(s,t)\right\} ,\nonumber\\
B^+_{jk}(s,t)&=&-\frac{2h}{9F^2\clm_\Delta^3(\clm_\Delta^2-s)}\left\{-f_1\frac{s-m_N^2-M_\pi^2}{2m_N}\mathcal{G}_B(s,t)\right.\nonumber\\
&&\hspace{1cm}\left.+\left[-f_2\frac{(s-m_N^2-M_\pi)^2}{4m_N^2}+2(2f_4-f_5)M_\pi^2\right]\mathcal{F}_B(s,t)\right\}  ,
\nonumber\\
 A^-_{jk}(s,t)&=&-\frac{1}{2}A^+_{jk}(s,t) , \qquad B^-_{jk}(s,t)=-\frac{1}{2}B^+_{jk}(s,t) .
\eea
\item{Born-terms of $\mathcal{O}(p^3)$} diagram (l):
\bea
A^+_l(s,t)&=&\frac{1}{9F^2\clm_\Delta^3(\clm_\Delta^2-s)}\left\{{b_3^2\mathcal{H}_A(s,t)}+2b_3b_8\frac{s-m_N^2-M_\pi^2}{2m_N}\mathcal{G}_A(s,t)\right.\nonumber\\
&&\hspace{1cm}\left.+b_8^2\frac{(s-m_N^2-M_\pi^2)^2}{4m_N^2}\mathcal{F}_A(s,t)\right\} ,\nonumber\\
B^+_l(s,t)&=&-\frac{h^2}{9F^2\clm_\Delta^3(\clm_\Delta^2-s)}\left\{{b_3^2\mathcal{H}_B(s,t)}+2b_3b_8\frac{s-m_N^2-M_\pi^2}{2m_N}\mathcal{G}_B(s,t)\right.\nonumber\\
&&\hspace{1cm}\left. +b_8^2\frac{(s-m_N^2-M_\pi^2)^2}{4m_N^2}\mathcal{F}_B(s,t)\right\} , \nonumber\\
 A^-_l(s,t)&=&-\frac{1}{2}A^+_l(s,t) , \qquad B^-_l(s,t)=-\frac{1}{2}B^+_l(s,t) .
\eea
\end{itemize}
Here the ${\cal F}$, ${\cal G}$ and ${\cal H}$ functions are defined as
\bea
\mathcal{F}_A(s,t)&=&(m_N+\clm_\Delta)\clm_\Delta^2\bigg[2(s-m_N^2)+3(t-2M_\pi^2)\bigg]\nonumber\\
&&+(s-m_N^2+M_\pi^2)\bigg[(s-m_N^2+M_\pi^2)m_N+2\clm_\Delta M_\pi^2\bigg], \nonumber\\
\mathcal{F}_B(s,t)&=&\clm_\Delta^2\bigg[4m_N(m_N+\clm_\Delta)+(4M_\pi^2-3t)\bigg]\nonumber\\
&&+(s-m_N^2+M_\pi^2)\bigg[(m_N^2-s-M_\pi^2)-2m_N\clm_\Delta\bigg],
\eea
\bea
\mathcal{G}_A(s,t)&=&m_N\clm_\Delta\bigg[(s-m_N^2)^2-M_\pi^2\bigg]+(s-m_N^2)(s-m_N^2+M_\pi^2)^2\nonumber\\
&&+\clm_\Delta^2\bigg[(s-m_N^2)^2+3(s-m_N^2)(t-2M_\pi^2)+M_\pi^4\bigg], \nonumber\\
\mathcal{G}_B(s,t)&=&-\clm_\Delta\bigg[(s+M_\pi^2)^2-4m_N^2(s+M_\pi^2)+3M_N^4\bigg]+(4M_\pi^2-3t)\clm_\Delta^3\nonumber\\
&&+m_N(s-m_N^2+M_\pi^2)^2+m_N\clm_\Delta^2\bigg[2(s-m_N^2)+3(t-2M_\pi^2)\bigg],
\eea
\bea
\mathcal{H}_A(s,t)&=&2\clm_\Delta(s-m_N^2)^2(s-m_N^2+M_\pi^2)-m_N(s-m_N^2)(s-m_N^2+M_\pi^2)^2\nonumber\\
&&+\clm_\Delta^2(\clm_\Delta-m_N)\bigg[3(s-m_N^2)(t-2M_\pi^2)+2M_\pi^2\bigg], \nonumber\\
\mathcal{H}_B(s,t)&=&2m_N\clm_\Delta[2(s-m_N^2)+M_\pi^2](m_N^2-s-M_\pi^2)+6m_N\clm_\Delta^3(2M_\pi^2-t)\nonumber\\
&&+\clm^2\bigg[3t(s+m_N^2)-4M_\pi^2(s+2m_N^2)\bigg]+(s+m_N^2)(s-m_N^2+M_\pi^2)^2.
\eea

\section{Redefinition of the LECs}
For simplicity, the following abbreviations are used:
\bea
\Sigma_{23}&=&\mn+\md , \qquad \Delta_{23}=\mn-\md , \qquad
\mathcal{Y}(a,b)\equiv2 a \mn+b \Sigma_{23} , \nonumber\\
\Delta_{(b_3,b_8,f_1,f_2)} & \equiv & 2\mn\mathcal{Y}(b_3,b_8)+\Delta_{23}\Sigma_{23}\mathcal{Y}(f_1,f_2)  .
\eea
In order to absorb the non-pole parts of the contributions of the $\mathcal{O}(p^2)$ and $\mathcal{O}(p^3)$ order delta-exchange diagrams,
the following redefinition of the LECs in the contact terms are needed
\bea
c_i \to c_i+\delta c_i , \qquad d_j \to d_j+\delta d_j  ,
\eea
where the shifts of the $c_i$ have the form
\bea
\delta c_1&=&0 \,, \quad
\delta c_2= \frac{1}{9\md^2}\bigg\{2h\,\Delta_{(b_3,b_8,f_1,f_2)}+\frac{\Delta_{23}}{4\,\mn^2}\Delta_{(b_3,b_8,f_1,f_2)}^2\bigg\}\,,\nonumber\\
\delta c_3&=&\frac{\md^2}{\mn^2}\delta c_2 \,, \qquad
\delta c_4=\frac{\md^2}{2\mn^2}\delta c_2 \,,
\end{eqnarray}
and the shifts for the $d_j$ read
\bea
\delta d_L&=&\frac{1}{18\mn\md}\bigg\{2h\left[2b_8\mn+\Sigma_{23}\mathcal{Y}(f_1,f_2)\right]-\left[\mathcal{Y}(b_3,b_8)-2b_8\Delta_{23}\right]\mathcal{Y}(b_3,b_8)\bigg\} ,\nonumber\\
\delta d_3&=&\frac{1}{36\md^2}\bigg\{2h\left[-2b_8\mn+\Sigma_{23}\mathcal{Y}(f_1,f_2)\right]+\left[\mathcal{Y}(b_3,b_8)-4(b_3+b_8)\mn\right]\mathcal{Y}(b_3,b_8)\bigg\} ,\nonumber\\
\delta d_T&=&-\frac{h}{36\mn^2\md^2}\Biggl\{2\mn[2\mn\mathcal{Y}(b_3,b_8)-b_8\md(\Sigma_{23}+2\md)]\nonumber\\
&+& (2\mn^3-\mn^2\md+3\md^3)\mathcal{Y}(f_1,f_2)\Biggr\} ,\nonumber\\
&-& \frac{1}{\mn^2\md^2}\left\{\mathcal{Y}(b_3,b_8)\bigg[(2\Delta_{23}-\md)\Sigma_{23}\mathcal{Y}(b_3,b_8)-2b_8\md(\Sigma_{23}+2\md)\Delta_{23}\bigg]\right\} ,\nonumber\\
\delta d_5&=&\frac{h}{72\mn^2\md^2}\left\{-8b_3\mn^3-\Sigma_{23}(\mn+\Sigma_{23})\left[2b_8\mn-\Delta_{23}\mathcal{Y}(f_1,f_2)\right]\right\}\nonumber\\
&&-\frac{1}{144\mn^2\md^2}\Delta_{23}\mathcal{Y}(b_3,b_8)\left\{(\mn+\Delta_{23}) \mathcal{Y}(b_3,b_8)+2b_8\md\Sigma_{23}\right\},
\eea
where $d_L\equiv d_{14}-d_{15}$ and $d_T\equiv d_1+d_2$.

\section{Definitions of the one-loop integrals}
Using notations similar to Ref.~\cite{Denner:2005nn}, 
the one-loop $n$-point integrals are defined by
\bea
\mH^{\mu_1\cdots\mu_P}_{a_1\cdots a_n} &=& \frac{(2\pi\mu)^{4-d}}{i\pi^2} \label{loopintdef}\\
&\times&  \int \frac{{\rm d}^d k \, k_{\mu_1}\cdots k_{\mu_P}}{\left[k^2-m_{a_1}^2+i\epsilon\right]\left[(k+p_1)^2-m_{a_2}^2
+i\epsilon\right]\cdots \left[(k+p_{n-1})^2-m_{a_n}^2+i\epsilon\right]},
\nonumber
\eea
with $a_j\in\{\pi, N,\Delta\}$, $j=1,\cdots, n$.
The results of integrals can be written in terms of the external momenta $p_i$ 
as (we need up to 4-point functions)
\bea\label{eq.ABCD}
&&\mH^{\mu_1\cdots\mu_P}_{a_1} ,  \quad\mH^{\mu_1\cdots\mu_P}_{a_1a_2}(p_1^2) \, ,\quad\mH^{\mu_1\cdots\mu_P}_{a_1a_2a_3}(p_1^2,(p_2-p_1)^2,p_2^2)\, ,\nonumber\\
&&\mH_{a_1a_2a_3a_4}^{\mu_1\cdots\mu_P}(p_1^2,(p_2-p_1)^2,(p_3-p_2)^2,p_3^2,p_2^2,(p_3-p_1)^2)\, .
\eea
Scalar integrals correspond to $P=0$ and for the tensor integrals $P\neq 0$.

The Passarino-Veltman decomposition expresses the tensor integrals in terms of Lorentz structures depending on the metric tensors and external momenta,
for example for the one-point functions we have
\bea
\mH^\mu_{a_1}=0\, , \quad \mH^{\mu\nu}_{a_1}=g^{\mu\nu}\mH^{(00)}_{a_1}\,,\cdots,
\eea
and for 2-point functions
\bea
\mH_{a_1a_2}^\mu&=&p_1^\mu \mH_{a_1a_2}^{(1)} , \quad \mH_{a_1a_2}^{\mu\nu}=g^{\mu\nu}\mH_{a_1a_2}^{(00)}+p_1^\mu p_1^\nu \mH_{a_1a_2}^{(11)} , \nonumber\\
 \mH_{a_1a_2}^{\mu\nu\rho}&=&(g^{\mu\nu}p_1^\rho+g^{\mu\rho}p_1^\nu+g^{\nu\rho}p_1^\mu)\mH_{a_1a_2}^{(001)}
+p_1^\mu p_1^\nu p_1^\rho \mH_{a_1a_2}^{(111)} ,\cdots.
\eea
The above decompositions are needed for the self-energies in the following section, and we refer the reader to Ref.~\cite{Denner:2005nn} for the higher rank tensors and more-point functions.

We denote loop integrals with removed UV-divergent parts (multiples of $R$) by $\bar{\mH}$ and loop integrals in chiral limit  (i.e. for $M^2\to 0$) without divergent pieces are labelled by $\mHO$. For example,
\bea
\mHO_\pi=0\, ,\qquad \mHO_{\pi N}(\mn^2)=\left\{\bar{\mH}_{\pi N}(\mn^2)\right\}_{M^2\to 0}  .
\eea

\section{\label{sec.SE} Self-energies of the nucleon and the delta}

The self-energy of the nucleon at leading one-loop order reads
\bea
\Sigma_A^{N,{\rm loop}}(s)&=& \frac{3 g^2}{4\fp^2}\left\{\mH_N+\mpi^2\mH_{\pi N}(s)+(s-\mn^2)\mH^{(1)}_{\pi N}(s)\right\}+\frac{(d-2)\ha^2}{2(d-1)\fp^2\md^2}\times\nonumber\\
&&\left\{(s-\md^2-3\mpi^2)\mH_\pi-2(\mH_\pi^{(00)}-\mH^{(00)}_\Delta)+\lambda_{\pi\Delta}(s)\left[\mH_{\pi \Delta}(s)+\mH^{(1)}_{\pi \Delta}(s)\right]\right\} ,\nonumber\\
\Sigma_B^{N,{\rm loop}}(s)&=&\frac{3g^2\mn}{4\fp^2}\left\{\mH_N+\mpi^2\mH_{\pi N}(s)\right\}+\frac{(d-2)\ha^2}{2(d-1)\fp^2\md}\left\{(s-\md^2-3\mpi^2)\mH_\pi\right.\nonumber\\
&&\left.+(s-\mpi^2+\md^2)\mH_\Delta-\lambda_{\pi\Delta}(s)\mH_{\pi\Delta}(s)\right\} ,
\eea
and the self-energy of the delta is
\bea
\Sigma_1^\Delta(s)&=&-\frac{\ha^2\mn}{\fp^2}\mH^{(00)}_{\pi N}(s)+\frac{5g_1^2}{12(d-1)\fp^2\md}
\left\{ -(d-1)\md^2 \left[\mH_\Delta+\mpi^2\mH_{\pi\Delta}(s)\right]- \right.\nonumber\\
&&\left.+(d-2)\left[\mH_\Delta^{(00)}+\mpi^2\mH^{(00)}_{\pi\Delta}(s)\right]+2\left[\mH_\pi^{(00)}+(s+\md^2)\mH^{(00)}_{\pi\Delta}(s)\right]\right\},\nonumber\\
\Sigma_6^\Delta(s)&=&-\frac{\ha^2}{\fp^2}\left\{\mH^{(00)}_{\pi N}(s)+\mH^{(001)}_{\pi N}(s)\right\}
+\frac{5g_1^2}{12(d-1)\fp^2\md^3}\left\{4\md^2\mH^{(00)}_{\pi\Delta}(s)\right.\nonumber\\
&&-(d-1)\md^2\left[\mH_\Delta+\mpi^2\mH_{\pi \Delta}(s)+(s-\md^2)\mH_{\pi \Delta}^{(1)}(s)\right]\nonumber\\
&&\left.+(d-2)\left[\mH_\Delta^{(00)}+\mpi^2\mH^{(00)}_{\pi\Delta}(s)+(s-\md^2)\mH_{\pi\Delta}^{(001)}(s)\right]\right\}.
\eea

\section{Counter terms in the EOMS scheme}
\label{deltaX}

In general, all LECs generate counter terms in the EOMS scheme as follows:
\bea
X&=&X_R+\frac{\bar{\delta}X}{16\pi^2F^2}R+\frac{\bar{\bar\delta}X}{16\pi^2F^2}\ ,\qquad X\in\{\ga,\ha,\mn,\md,a_1,c_{i=1,\cdots, 4}\} \, ,\nonumber\\
Y&=&Y_R+\frac{\bar{\delta}Y}{16\pi^2F^2}R\, ,\quad Y\in\{\ell_3,\ell_4,d_{1}+d_2,d_3,d_5,d_{14}-d_{15},d_{18}-2d_{16}\} \, .
\eea
%The LECs in Set X need to be performed first UV-infinite shift and then PCBTs-finite shift, while the ones in Set Y only UV-infinite shift.
We have derived all counter terms explicitly and most of them turn out to be too lengthy to be shown here.
Hence, we only show the counter terms for the parameters involved in the nucleon and delta mass renormalization.

The infinite parts of counter terms for $\mn$, $\md$, $c_1$ and $a_1$ are
\bea
{\bar \delta}\mn&=&-\frac{\mn_R^2 \left(\ha_R^2 \left(\mn_R^3+2 \mn_R^2 m_{\Delta R}-2 \mn_R m_{\Delta R}^2-6 m_{\Delta R}^3\right)
-9 \ga_R^2 \mn_R m_{\Delta R}^2\right)}{6 m_{\Delta R}^2}\, ,\\
%\eea
%
%\bea
{\bar\delta}\md&=&\frac{110 g_{1R}^2 m_{\Delta R}^3+9 \ha_R^2 \left(6 \mn_R^3+2 \mn_R^2 m_{\Delta R}-2 \mn_R m_{\Delta R}^2-m_{\Delta R}^3\right)}{216 }\, ,
\\
{\bar \delta}c_1&=&-\frac{\mn _R\left(9 \ga_R^2 m_{\Delta R}^2+2 \ha_R^2 \mn_R (2 \mn_R+3 m_{\Delta R})\right)}{24 m_{\Delta R}^2}\, ,\\
%\eea
%
%\bea
{\bar \delta}a_1&=&-\frac{50 g_{1R}^2 m_{\Delta R}+9 \ha_R^2 (3 \mn_R+2 m_{\Delta R})}{432} \, .
\eea
The finite parts of counter terms for $\mn$, $\md$, $c_1$ and $a_1$ are
\bea
\bar{\bar\delta}\mn&=&\frac{(2-d) \ha_R^2 \mHO_\Delta \left(d \,\mn_R^4+2 d \,\mn_R^3 m_{\Delta R}-2 (d-2) \mn_R^2 m_{\Delta R}^2+2 d\, \mn_R m_{\Delta R}^3
+d \,m_{\Delta R}^4\right)}{4 (d-1) d \, \mn_R m_{\Delta R}^2}\nonumber\\
&&-\frac{3 \ga_R^2 \mn_R \mHO_N}{2 }+\frac{(d-2) \ha_R^2 (\mn_R-m_{\Delta R})^2 (\mn_R+m_{\Delta R})^4 \mHO_{\pi\Delta}(\mn_R^2)}{4 (d-1)  \mn_R m_{\Delta R}^2}\, ,\\
%\eea
%
%\bea
\bar{\bar\delta}\md&=&\frac{\ha_R^2 (\mn_R-m_{\Delta R})^2 (\mn_R+m_{\Delta R})^4 \mHO_{\pi N}(m_{\Delta R}^2)}{8 (d-1)  m_{\Delta R}^3}
-\frac{5 ((d-2) (d-1) d-2) g_{1R}^2 m_{\Delta R} \mHO_\Delta}{6 (d-1)^2 d }\nonumber\\
&&-\frac{\ha_R^2 \mHO_N \left(d \,\mn_R^4+2 d\, \mn_R^3 m_{\Delta R}-2 (d-2) \mn_R^2 m_{\Delta R}^2+2 d \,\mn_R m_{\Delta R}^3+d\, m_{\Delta R}^4\right)}{8 (d-1) d\,  m_{\Delta R}^3}\, ,
\eea

\bea
\bar{\bar\delta}c_1&=& \frac{(d-2) h_R^2 (\mn_R+m_{\Delta R})^2 \left(d \left(\mn_R^2+m_{\Delta R}^2\right)-2 \ga_R \mn_R\right){ \mHO}_{\pi\Delta}(\mn_R^2)}{16 (d-1)
\mn_R m_{\Delta R}^2}\nonumber\\
&&+\frac{3 (d-2) \ga_R^2 \mHO_N}{16 (d-3) \mn_R}-\frac{(d-2) h_R^2 \mHO_\Delta\left(d (\mn_R+m_{\Delta R})^2-2 \mn_R m_{\Delta R}\right)}{16 (d-1)  \mn_R m_{\Delta R}^2}\, ,\label{eq.EOMSc1}\\
%\eea
%
%\bea
\bar{\bar\delta}a_1&=&\frac{5 ((d-2) d+2) g_{1R}^2 \mHO_\Delta}{48 (d-1)^2 F^2 m_{\Delta R}}-\frac{\ha_R^2 \mHO_N \left(d (\mn_R+m_{\Delta R})^2-2 \mn_R m_{\Delta R}\right)}{32 (d-1) F^2 m_{\Delta R}^3}\nonumber\\
&&+\frac{\ha_R^2 (\mn_R+m_{\Delta R})^2 \left(d \left(\mn_R^2+m_{\Delta R}^2\right)-2 \mn_R m_{\Delta R}\right) \mHO_{\pi N}(m_{\Delta R}^2)}{32 (d-1) F^2 m_{\Delta R}^3}\, .
\eea

\section{One loop contributions in the baryon octet self energy}
\label{OSelfEnergy}

We use the definitions and notations of Ref.~ \cite{Lehnhart:2004vi} and the LO meson-octet-decuplet interaction term is taken from
Ref.~\cite{Borasoy:1996bx} with the coupling constant ${\mathscr C}$.  Contact interaction and one loop diagrams contributing to the octet masses are shown in Fig.~\ref{fig:m} a), where the solid, dashed and double lines correspond to the octet brayons, mesons and the decuplet baryons, respectively.

The contributions of NLO contact interactions to the octet baryon masses can be found e.g. in Ref.~ \cite{Lehnhart:2004vi} and the one loop order
contributions to the octet self energy are specified below.  $\Sigma^{ab }_{oct} $ corresponds to the diagram with octet baryon propagators in the loop and $\Sigma^{ab }_{dec}$ to the one
with decuplet baryon propagators. Summation over repeated indices is implied.
\begin{eqnarray}
\Sigma^{ab }_{oct} &=& \frac{i m_B}{8 \pi ^2 F_\phi^2}  \left(M_d^2
   B_0\left(m_B^2,m_B^2,M_d^2\right)+
   A_0\left(m_B^2\right)\right) (D d^{dca}+i F f^{dca})
   (F f^{dbc}-i D d^{dbc}),\nonumber\\
\label{OSEO}
\Sigma^{ab }_{dec} &=& \left[{\rm Tr}\left\{\lambda^a \lambda^b \right\} {\rm Tr}\left\{\lambda^d \lambda^d \right\}
-{\rm Tr}\left\{\lambda^a \lambda^d \lambda^d \lambda^b \right\} \right] \nonumber\\
&\times&
\Biggl\{-\frac{{\mathscr C}^2
   \left((m_B-m_D)^2-M_d^2\right)
   \left((m_B+m_D)^2-M_d^2\right)^2
   B_0\left(m_B^2,m_D^2,M_d^2\right)}
   {1536 \pi ^2 F_\phi^2 m_B
   m_D^2}\nonumber\\
  &+& \frac{{\mathscr C}^2
   A_0\left(m_D^2\right)
   \left(-M_d^2+m_B^2-m_B
   m_D+m_D^2\right)
   \left(-M_d^2+m_B^2+3 m_B
   m_D+m_D^2\right)}{1536 \pi ^2 F_\phi^2 m_B m_D^2}\nonumber\\
   &+& \frac{{\mathscr C}^2
   A_0\left(M_d^2\right) \left(M_d^2 \left(3
   m_B^2+2 m_B m_D+2
   m_D^2\right)-M_d^4+(m_B-m_D)
   (m_B+m_D)^3\right)}{1536 \pi ^2
   F_\phi^2 m_B
   m_D^2}\nonumber\\
   &+& \frac{m_B {\mathscr C}^2 }{9216 \pi ^2 F_\phi^2
   m_D^2} \biggl[-4
   m_B M_d^2 (2 m_B+3 m_D)-3 M_d^4+2
   m_B^4+4 m_B^3 m_D\nonumber\\
   &-& 4 m_B^2
   m_D^2-12 m_B m_D^3+3
   m_D^4\biggr] \Biggr\}.
\label{OSED}
\end{eqnarray}
We renormalize these loop contributions by applying the EOMS scheme without expanding in powers of $m_D-m_B$, i.e. we expand in powers of the meson masses and absorb terms of order zero in the renormalization of the mass in the chiral limit and the order two terms in the renormalization of contact interactions. We checked that this renormalization indeed can be carried out self-consistently.

%\newpage

\end{document}